\documentclass[12pt]{iopart}
\usepackage{iopams}
\usepackage{graphicx,bm} 	
	\usepackage{color}
\usepackage{hyperref}			
\usepackage{fancyhdr}			
\usepackage{microtype}		
\usepackage{placeins}			

\let\SQRT\sqrt
\renewcommand{\sqrt}[1]{\ensuremath{\SQRT{#1} \;}}
\catcode`_=\active
\newcommand_[1]{\ensuremath{\sb{\mathrm{#1}}}}

\newcommand{\de}[2]{\ensuremath{\frac{\textup{d} #1}{\textup{d} #2}}}   
\newcommand{\sde}[2]{\ensuremath{\frac{\textup{d}^{2} #1}{\textup{d} #2^{2}}}} 

\newcommand{\intinf}[1]{\ensuremath{\int\sb{-\infty}^{\infty}} \textup{d}#1 \;} 

\newcommand{\modd}[1]{\ensuremath{\left| #1 \right|}} 
\newcommand{\td}{\ensuremath{\left(t\right)}} 
\newcommand{\fd}{\ensuremath{\left(\omega\right)}} 

\newcommand{\Br}[1]{\ensuremath{\left( #1 \right)}} 
\newcommand{\Sq}[1]{\ensuremath{\left[ #1 \right]}} 
\newcommand{\Cu}[1]{\ensuremath{\left\{ #1 \right\}}} 

\newcommand{\expect}[1]{\ensuremath{\left< #1 \right>}}
\newcommand{\comm}[2]{\ensuremath{\left[ #1,#2 \right]}}
\newcommand{\p}{\ensuremath{\hat{p}}}   
\newcommand{\x}{\ensuremath{\hat{x}}}   
\newcommand{\baop}[1]{\ensuremath{\hat{#1}}}   
\newcommand{\bcop}[1]{\ensuremath{\hat{#1}^\dagger}}   

\newcommand{\SNR}{\ensuremath{\textup{SNR}}} 

\newcommand{\half}{\ensuremath{\frac{1}{2}}}   
\newcommand{\E}{\mathrm{e}} 
\newcommand{\I}{\mathrm{i}} 
\newcommand{\isotope}[2]{\textsuperscript{#2}#1} 

\newcommand{\FP}{Fabry-P\'{e}rot } 
\newcommand{\via}{via} 
\newcommand{\ie}{\textit{i.e.}} 
\newcommand{\eg}{\textit{e.g.}} 
\newcommand{\etc}{\textit{etc.}} 
\newcommand{\viz}{\textit{viz}.} 
\newcommand{\cf}{\textit{cf}.} 

\newcommand{\eqref}[1]{(\ref{#1})} 

\begin{document}

\title[Coherent control and feedback cooling in an atom--optomechanical system]{Coherent control and feedback cooling in a remotely-coupled hybrid atom--optomechanical system}
\author{James S. Bennett, Lars S. Madsen, Mark Baker, Halina Rubinsztein-Dunlop and Warwick P. Bowen}
\address{Australian Research Council Centre of Excellence for Engineered Quantum Systems (EQuS), The University of Queensland, St Lucia, QLD 4072, Australia}
\ead{james.bennett2@uqconnect.edu.au}

\begin{abstract}
Cooling to the motional ground state is an important first step in the preparation of nonclassical states of mesoscopic mechanical oscillators. Light-mediated coupling to a remote atomic ensemble has been proposed as a method to reach the ground state for low frequency oscillators. The ground state can also be reached using optical measurement followed by feedback control. Here we investigate the possibility of enhanced cooling by combining these two approaches. The combination, in general, outperforms either individual technique, though atomic ensemble-based cooling and feedback cooling each individually dominate over large regions of parameter space. 
\end{abstract}

\maketitle


\section{Introduction} \label{Sec:Introduction}

Preparation of mesoscopic mechanical devices in high-purity nonclassical states is a long-standing goal of the opto- and electromechanical communities. In addition to promising applications in fundamental physics research---such as gravitational effects in quantum mechanics and the quantum-to-classical transition \cite{Treutlein2012,Chen2013}---mechanical devices provide outstanding opportunities for metrology \cite{Giovannetti2004,Milburn2011,Serafini2012} and emerging quantum technologies \cite{Tsukanov2011,Muschik2011,Stannigel2012}.

The majority of mechanical quantum state preparation and verification schemes require that the oscillator be initialised very near to the motional ground state \cite{Kleckner2008,Rocheleau2010}. Though this has been achieved in the gigahertz and megahertz regimes \cite{OConnell2010,Teufel2011,Chan2011}, progress toward cooling low-frequency ($\omega_{m}$) oscillators has been inhibited by the lack of both cryogenic systems capable of ‘freezing out’ their thermal motion and sufficiently high quality optical cavities to achieve the good cavity (cavity linewidth $\kappa \ll \omega_{m}$) regime required for resolved-sideband cooling.

Both remote coupling to the motional state of a cooled atomic ensemble \cite{Hammerer2010, Camerer2011, Vogell2013} and optical measurement followed by feedback control \cite{Mancini1998,Courty2001,Vitali2002,Hopkins2003,Genes2008a, Doherty2012} have been suggested as alternative approaches to cooling that, in principle, overcome these limitations and allow ground state cooling in the bad cavity limit ($\kappa \gg \omega_{m}$).

In remote atomic ensemble-based cooling, or \textit{sympathetic cooling}, light mediates a swap between the centre of mass motional state of the ensemble and a mode of the mechanical oscillator, which may be separated from the former by a macroscopic distance ($\sim 1$ m) \cite{Hammerer2010,Camerer2011,Vogell2013}. Proposed remotely-coupled atom--optomechanical systems are theoretically capable of sympathetically cooling their mechanical elements to near the ground state. The directional flow of quantum information from the mechanical element---the `plant', in control parlance---to the atoms and back allows us to identify the atoms as an irreversible coherent controller (\cf{} \cite{Jacobs2014}, pg. 3, and \cite{Lloyd2000,Hamerly2012}). The controller is necessarily imperfect, as the phase of the output optical field retains knowledge of the mechanical position. The resulting back-action on the momentum hinders the cooling process.

\begin{figure}[bt]
\centering
\def\svgwidth{1\columnwidth}
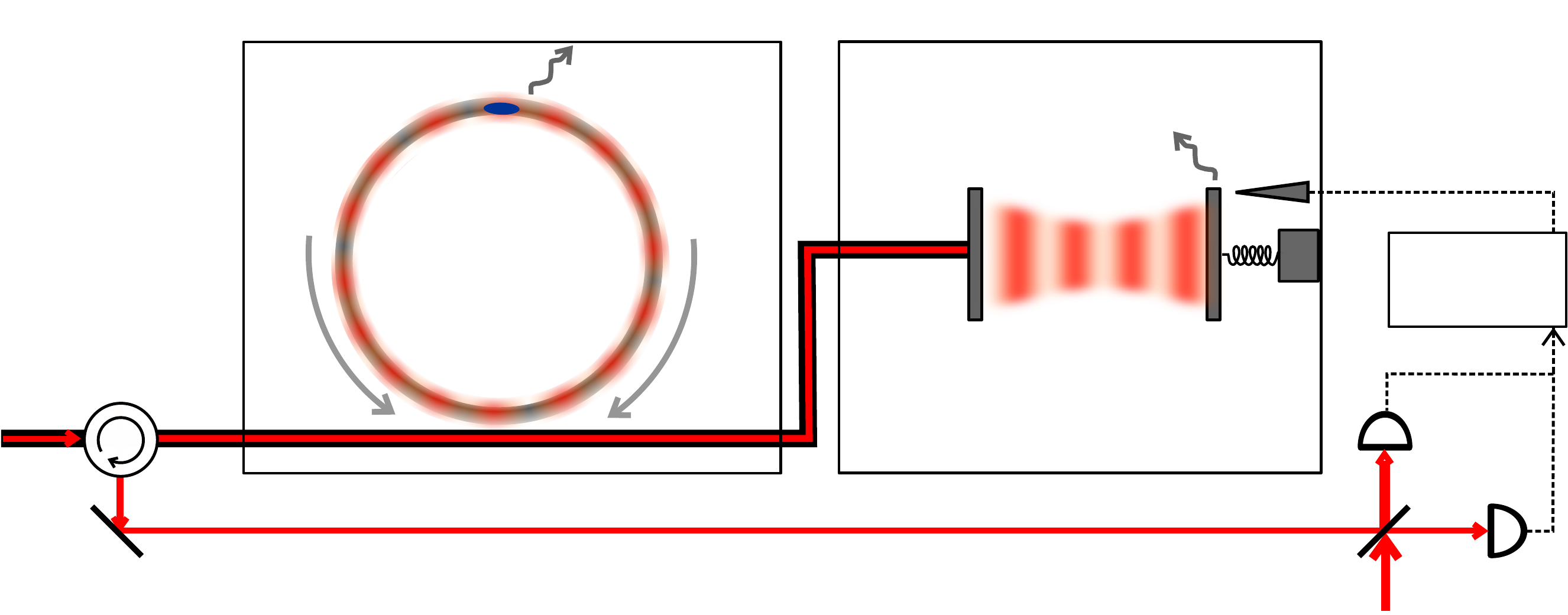
\caption{\label{Fig:ModelGeometry} A generic remotely-coupled hybrid atom--optomechanical system incorporating continuous measurement and feedback. The counter-propagating optical modes of a ring cavity (AC) are coupled to the motion of an atomic ensemble; similarly, the modes a single-sided cavity (MC) interact with a micromechanical oscillator. The transfer of light between these two subsystems couples the collective motion of the atoms to that of the mechanical element. Light exiting the system may be mixed with a bright local oscillator (LO) in order to measure the (optical phase quadrature) mechanical position, allowing the application of a classical feedback force.}
\end{figure}

In this article we investigate combining sympathetic cooling with feedback damping based on a phase measurement of the output field (\cf{} Fig.~\ref{Fig:ModelGeometry}). The latter retrieves information from the optical field and allows suppression of decoherence. We derive an analytical mechanical power spectrum for such a system, from which the steady-state temperature is found. This reveals a set of criteria specifying when near-ground-state temperatures may be achieved. In general the combined scheme outperforms both individual methods; however, when the optomechanical cooperativity is sufficiently large the cooling is dominated by feedback. Conversely, one may still approach the mechanical ground state with weak optomechanical cooperativity provided that its ensemble--light counterpart (\cf{} Eqn~\eqref{Eqn:AtomCooperativity}) is appropriately large. These statements are made quantitative in $\S$~\ref{Sec:AtomsAlone}. We also clarify the role of atom--light and optomechanical interactions in performing the atom $\leftrightarrow$ mechanical state swap which underpins sympathetic cooling. In particular, and somewhat counter-intuitively, this swap is possible even when the optomechanical cooperativity is insufficient to permit a complete state transfer between the oscillator and the optical field.

\section{Model} \label{Sec:Theory}

Consider the device depicted in Fig.~\ref{Fig:ModelGeometry}, in which a ring resonator, containing an atomic cloud, and a one-sided optomechanical cavity are coupled by a lossless optical transmission line. An optical lattice, formed by interference of the counter-propagating cavity modes, traps the atoms in approximately harmonic wells (\cf{} $\S$~\ref{Sec:AtomLightInteraction}). The atomic centre of mass and the optically-coupled oscillator therefore comprise two harmonic mechanical degrees of freedom.

This class of hybrid atom--optomechanical system (\eg{} \cite{Hammerer2010,Camerer2011,Vogell2013} and \cite{Hammerer2009EPR}) is desirable from an immediate experimental perspective, where their key advantage is circumventing the need for close integration of cryogenic and ultra-high vacuum apparatus, and within the context of future quantum networks, where atomic and solid-state processing and memory nodes are anticipated to be interfaced \via{} optical photons \cite{Kimble2008}.

Displacement of the atoms relative to the lattice results in an exchange of photons between the left- and right-going optical modes, which modulates the optical power incident upon the mechanical device. Conversely, changing the position of the latter alters the phase of the reflected field (\cf{} $\S$~\ref{Sec:OptomechanicalInteraction}), causing axial translation of the optical lattice. In this way each oscillator is subject to a force which depends on the position of its counterpart \cite{Hammerer2010}.

Our model of the atom--mechanical coupling is closely related to that proposed by \cite{Vogell2013}. The primary difference, besides the inclusion of the external measurement-based feedback loop, is the addition of a second optical cavity into which the atoms are loaded. The cavity allows us to employ an intuitive two-mode description of the atom--light interaction and may be experimentally desirable in certain circumstances (\cf{} $\S$~\ref{Sec:EffectiveDynamics}).

\subsection{\textbf{Atom--light interaction}} \label{Sec:AtomLightInteraction}

The heart of the atom trapping apparatus is an optical dipole trap \cite{Grimm2000} which uses the AC Stark effect to confine $N$ identical two-level atoms of mass $m$. This trap is incorporated into a ring resonator of quality $Q_{AC}$ which is driven by an input field $\baop{a}_{d}$, with a large coherent amplitude $\alpha_{d} = \expect{\baop{a}_{d}}$ at (angular) frequency $\omega_{d}$. Neglecting internal losses, the linewidth of this cavity is $\kappa_{AC} \approx \omega_{d}/Q_{AC}$ (\ie{} the resonator is strongly overcoupled to the transmission line). We take the majority of the cavity's modes to be far off resonance with the drive beam, leaving two relevant counter-propagating modes which are lowered by the operators $\baop{a}_{L}$ and $\baop{a}_{R}$, with the subscripts referring to the handedness of circulation depicted in Fig.~\ref{Fig:ModelGeometry}.

In practice additional field modes or semiclassical potentials may be required to stably confine the atoms in the plane transverse to the counter-propagating modes introduced above. We will neglect all noise introduced by these fields/potentials in the analysis that follows on the grounds that, to harmonic order, motion in the transverse directions is decoupled from the axial motion.

The Hamiltonian governing the internal (axial) dynamics of the cavity---in a frame rotating at the drive beam frequency $\omega_{d}$---is then
\begin{equation*}
\hat{H}_{AC} = -\hbar\Delta_{AC}\Br{\bcop{a}_{R}\baop{a}_{R} + \bcop{a}_{L}\baop{a}_{L}} +\hat{H}_{SS} +\sum\sb{j=1}^{N} \frac{\p\sb{j}^{2}}{2m},
\end{equation*}
where $\Delta_{AC}$ is the detuning of the drive from the bare cavity resonance, $\p\sb{j}$ is the momentum of the $j$\textsuperscript{th} atom and $\hat{H}_{SS}$, the Stark shift operator, models the light--atom interactions.

Following \cite{Hammerer2010} (\cf{} \cite{Dalibard1985}), we will treat all atom--light interactions in the dispersive limit, suppressing any internal structure of the ground and first excited electronic states of the atom. This is a valid approximation for alkali gases provided that the detuning $\Delta_{t}$ between the drive and the electronic transition frequency is large compared to the laser linewidth and all other relevant frequency scales \cite{Grimm2000}. The one-dimensional Hamiltonian describing the atom--light interaction is therefore (neglecting off-resonant terms)
\begin{equation}
\hat{H}_{SS} =  \sum\sb{j=1}^{N} \frac{\mu^{2}}{\hbar\Delta_{t}}\hat{E}^{\left(-\right)}\left(\x\sb{j}\right)\hat{E}^{\left(+\right)}\left(\x\sb{j}\right),
\label{Eqn:AtomLight}
\end{equation}
where $E^{\Br{+}}$ is the positive frequency component of the electric field and each atom has a transition dipole moment of $\mu$ \cite{Vogell2013}.

The positive-frequency part of the cavity electric field may be written
\begin{equation*}
\hat{E}^{\Br{+}} =\I\sqrt{\frac{\hbar\omega_{AC}}{2\epsilon_{0}\mathcal{V}}}\Br{\baop{a}_{R}\E^{-\I kx} + \baop{a}_{L}\E^{\I kx}},
\end{equation*}
where $\omega_{AC}$ is the bare resonance frequency of the atom cavity, $k$ is the optical wavenumber, $\epsilon_{0}$ is the permittivity of free space and $\mathcal{V}$ is the cavity mode volume \cite{WallsMilburn2008}. From this we find \via{} Eqn~\eqref{Eqn:AtomLight} that the strength of the single-atom--light interaction is characterised by the coupling rate
\begin{equation}
g_{a} = \frac{\mu^{2}\omega_{d}}{2 \hbar \Delta_{t} \epsilon_{0} \mathcal{V}}.
\label{Eqn:AtomCouplingRate}
\end{equation}
We have used $\omega_{AC} \approx \omega_{d}$. Note that for red detuned light ($\Delta_t<0$), as we will assume from here onward, $g_{a}$ is negative.

Expanding the annihilation operators $\baop{a}_{L,R} \rightarrow \alpha_{L,R} + \delta \baop{a}_{L,R}$ about the coherent field amplitudes $\expect{\baop{a}_{L,R}} = \alpha_{L,R}$, which we assume are real and satisfy $\alpha_{R} \approx \alpha_{L} \gg 1$, and truncating the oscillatory terms at second order in the Lamb-Dicke parameter \cite{Hammerer2010}, we acquire a static shift of the cavity resonance frequency, an effective harmonic trapping potential with frequency
\begin{equation}
\omega_{a} = 2k\sqrt{\frac{-2\hbar g_{a}\alpha_{L}\alpha_{R}}{m}} \nonumber
\end{equation}
and a linearised interaction between the atoms' positions and the optical phase quadrature fluctuations. Finally, setting the bare detuning to $\Delta_{AC} = -2Ng_{a}$ brings the cavity onto resonance in the presence of the mean interaction, yielding the Hamiltonian
\begin{equation}
\hat{H}_{AC} =  \sum\sb{j=1}^{N} \Sq{\frac{\p\sb{j}^{2}}{2m} + \frac{m\omega_{a}^{2}\x\sb{j}^{2}}{2} + 2\hbar kg_{a}\x\sb{j}\Br{\alpha_{L} \delta \hat{X}_{R}^{-}-\alpha_{R} \delta \hat{X}_{L}^{-}}}
\label{Eqn:AtomLightLin}
\end{equation}
where we have introduced the amplitude and phase quadrature fluctuation operators, $\delta \hat{X}^{+} = \delta\bcop{a} + \delta\baop{a}$ and $\delta \hat{X}^{-} = \I\left(\delta\bcop{a} - \delta\baop{a}\right)$ and neglected contributions of order $\x\sb{j}^{2}\delta \hat{X}^{\pm}$.

Inspection of Eqn~\eqref{Eqn:AtomLightLin} reveals that the phase difference of the optical fields couples to the \textit{collective} motion of the atomic cloud, specifically the centre of mass mode. This degree of freedom may be described by a simple harmonic oscillator with coordinate $\x_{a} = \frac{1}{N}\sum\sb{j=1}^{N} \x\sb{j}$, momentum $\p_{a} = \sum\sb{j=1}^{N} \p\sb{j}$ and zero-point extension $x_{zp,a} = \sqrt{\hbar/2Nm\omega_{a}}$ (\cf{} $\S$~\ref{Sec:EffectiveDynamics}).

\subsection{\textbf{Optomechanical interaction}} \label{Sec:OptomechanicalInteraction}

The canonical cavity optomechanical interaction is most easily understood in the context of a single-sided \FP cavity wherein the input mirror is fixed and the other is harmonically bound (\eg{} as depicted in Fig.~\ref{Fig:ModelGeometry}). Motion of the end mirror changes the cavity length, thereby altering the optical resonance frequency, which in turn modulates the number of photons in the cavity field. Finally, the photon number controls the radiation force experienced by the mirror. This interplay leads to the emergence of a rich variety of well-studied physics \cite{Kippenberg2007,KippenbergVahala2008} even at first-order in $\x_{m}$.

A linearly-coupled optomechanical system, whether actuated by radiation pressure or the gradient force \cite{Thourhout2010}, may be described by the parametric coupling Hamiltonian \cite{Law1995}
\begin{equation*}
\hat{H}_{MC} = \hbar\Br{g_{m}\x^{\prime}_{m}-\Delta_{MC}}\bcop{b}\baop{b} +\frac{\p_{m}^{2}}{2M} + \half M\omega_{m}^{2}\x^{\prime\;2}_{m}
\end{equation*}
in a frame rotating at $\omega_{d}$. The bare detuning of the laser drive to the cavity resonance is $\Delta_{MC}$, $\x_{m}^{\prime}$ is a suitable position coordinate of a vibrational mode with effective mass $M$ and frequency $\omega_{m}$, photons are removed from the cavity field by $\baop{b}$ and $g_{m}$ is the optomechanical coupling rate (with dimensions of s\textsuperscript{-1}m\textsuperscript{-1}).

As above, we assume that the internal optical mode is coupled to the drive beam at a rate $\kappa_{MC} \approx \omega_{d}/Q_{MC}$, where $Q_{MC}$ is the Q-factor of the optical resonator, leading to the build-up of a steady-state intracavity amplitude $\beta = 2\alpha_{d}/\sqrt{\kappa_{MC}}$. Linearising about this amplitude, introducing a zero-mean position coordinate $\x_{m}$ and selecting $\Delta_{MC}$ so as to bring the cavity onto resonance in the presence of the mean interaction\footnote{For a linear oscillator we may arrive at Eqn~\eqref{Eqn:OptoMechLin} by introducing $\x_{m} = \x_{m}^{\prime}-\expect{\x_{m}^{\prime}}$, with $\expect{\x_{m}^{\prime}} = \frac{\hbar g_{m}}{M\omega_{m}^{2}}\beta^{2}$, and detuning from the bare cavity resonance by $\Delta_{MC}=g_{m}\expect{\x_{m}^{\prime}}$. Alternatively, an external control force may be used to cancel the mean optical force, leaving $\Delta_{MC} = 0$.} yields the effective Hamiltonian
\begin{equation}
\hat{H}_{MC} = \hbar g_{m}\x_{m}\beta\delta \hat{Y}^{+} + \frac{\p_{m}^{2}}{2M} + \half M\omega_{m}^{2}\x_{m}^{2}, \label{Eqn:OptoMechLin}
\end{equation}
with the optical quadrature fluctuation operators $\delta \hat{Y}^{\pm}$.

The ground state variance of the mechanical resonator is $x_{zp,m}^{2} = \hbar/2M\omega_{m}$.

\subsection{\textbf{Effective dynamics}} \label{Sec:EffectiveDynamics}
Given Eqn~\eqref{Eqn:AtomLightLin} and Eqn~\eqref{Eqn:OptoMechLin} we may determine the dynamics of the system in the Heisenberg picture. Under free evolution, neglecting coupling to reservoirs for the moment,
\begin{eqnarray}
Nm\sde{\hat{x}_{a}}{t} & = & -Nm\omega_{a}^{2}\x_{a}  + 2\hbar Nkg_{a}\left(\alpha_{R}\delta \hat{X}_{L}^{-}-\alpha_{L}\delta \hat{X}_{R}^{-}\right),  \label{Eqn:MechAC} \\
M\sde{\hat{x}_{m}}{t} & = & -M\omega_{m}^{2}\x_{m}  - \hbar g_{m}\beta\delta \hat{Y}^{+}. \label{Eqn:MechMT}
\end{eqnarray}
These equations show, as expected, that the atom--light interaction depends on the optical phase fluctuations, whereas the optomechanical system responds to amplitude noise (\cf{} \cite{Camerer2011,Vogell2013}).

Dramatic simplifications may be made in the optical adiabatic limit, wherein the optical quadrature fluctuations are slaved to the positions of the mechanical and atomic elements. Conveniently, this is also the regime in which the most sensitive measurements of mechanical displacement are achieved \cite{Genes2008a}. We make the bad-cavity approximation (\cf{} \ref{Sec:Fields}) under the requirement that, for both optical cavities, $\kappa\gg\max\left\{\omega_{a}, \omega_{m}\right\}$. What emerges is an effective direct coupling between the two oscillators;
\begin{eqnarray}
Nm\sde{\hat{x}_{a}}{t} & = & -Nm\omega_{a}^{2}\x_{a} + K\x_{m}, \label{Eqn:MechACEffNoLoss}\\
M\sde{\hat{x}_{m}}{t} & = & -M\omega_{m}^{2}\x_{m} +\hat{F}_{BA,m} + K\x_{a}. \label{Eqn:MechMTEffNoLoss}
\end{eqnarray}
$\hat{F}_{BA,m} = -4\hbar g_{m}\alpha_{d}\delta\hat{X}_{d}^{+}/\kappa_{MC}$ is an optical back-action force (\cf{} $\S~$\ref{Sec:Langevin}) arising due to amplitude fluctuations of the input electromagnetic field ($\delta\hat{X}_{d}^{+} = \delta\bcop{a}_{d} + \delta\baop{a}_{d}$); for the case discussed here the back-action on the atoms is negligibly small (\cf{} \ref{Sec:Fields}). The spring constant $K$ quantifies, for the moment, the coupling strength.
\begin{equation}
K = -8^{2} \hbar Nkg_{a}g_{m}\frac{Q_{AC}Q_{MC}}{\omega_{d}^{2}}\alpha_{d}^{2}\label{Eqn:SpringConstant}
\end{equation}

In this same limit the optical field which exits the system carries the fluctuations
\begin{eqnarray}
\delta \hat{Z}^{+} & = & \delta \hat{X}_{d}^{+} \nonumber\\
\delta \hat{Z}^{-} & = & \delta \hat{X}_{d}^{-}-\frac{4g_{m}\beta}{\sqrt{\kappa_{MC}}}\x_{m}. \label{Eqn:OutputField}
\end{eqnarray}

It is natural to divide the optically-mediated interactions into coherent and `incoherent' processes. The latter is simply the back-action noise. Taken together, Eqn~\eqref{Eqn:MechACEffNoLoss} and Eqn~\eqref{Eqn:MechMTEffNoLoss} imply that the former may be described by an effective direct Hamiltonian interaction between the two oscillators,
\begin{eqnarray}
\hat{H}_{eff} & = & - \hbar g \frac{\x_{m}}{x_{zp,m}}\frac{\x_{a}}{x_{zp,a}} \nonumber\\
& = &  -\hbar g \Br{\bcop{a}_{a}\baop{a}_{m}+\bcop{a}_{m}\baop{a}_{a}+\bcop{a}_{m}\bcop{a}_{a}+\baop{a}_{m}\baop{a}_{a}}, \label{Eqn:EffectiveHamiltonian}
\end{eqnarray}
where the motional annihilation operators are $\baop{a}_{a}$ (atomic) and $\baop{a}_{m}$ (mechanical), and $g$ is the atom--mechanical coupling rate \cite{Hammerer2010, Vogell2013}. The latter is
\begin{equation}
g = \sqrt{N}\frac{g_{m}}{k}\frac{\omega_{a}}{\kappa_{MC}} \sqrt{\frac{m\omega_{a}}{M\omega_{m}}}
\label{Eqn:CouplingRate}
\end{equation}
in accordance with the relationship $g = \frac{K}{\hbar}x_{zp,a}x_{zp,m}$.

The combined system is stable if
\begin{equation}
g < \half \frac{1}{1/\omega_{a} + 1/\omega_{m}}; \label{Eqn:Instability}
\end{equation}
at higher coupling rates the harmonic potential experienced by the symmetric mode $\x_{a}+\x_{m}$ becomes inverted, anti-trapping this degree of freedom.

Note that if optical losses between (and/or within) the cavities are non-negligible the effective interaction of the oscillators becomes non-Hamiltonian, as discussed in detail by \cite{Hammerer2010} (see also \ref{Sec:Fields}).

Our Eqn~\eqref{Eqn:CouplingRate} agrees with that derived in \cite{Vogell2013}, which sports a free-space atom trap.

One may be tempted to increase $g$ by choosing $\omega_{a} > \omega_{m}$; however, for our purposes this is counterproductive. Consider Eqn~\eqref{Eqn:EffectiveHamiltonian} in the interaction picture with respect to the free mechanical Hamiltonian ($\hbar\omega_{a}\bcop{a}_{a}\baop{a}_{a} + \hbar\omega_{m}\bcop{a}_{m}\baop{a}_{m}$);
\begin{eqnarray*}
H_{eff} & = & -\hbar g \left\{a^{\dagger}_{a}a_{m}\E^{+\I\Br{\omega_{m}-\omega_{a}}t}+a^{\dagger}_{m}a_{a}\E^{-\I\Br{\omega_{m}-\omega_{a}}t}\right.\\
& & \phantom{-\hbar g g} \left. +a^{\dagger}_{m}a^{\dagger}_{a}\E^{-\I\Br{\omega_{m}+\omega_{a}}t}+a_{m}a_{a}\E^{+\I\Br{\omega_{m}+\omega_{a}}t}\right\},
\end{eqnarray*}
where the lack of a carat indicates an operator in the interaction picture. If the mechanical systems are not resonant ($\omega_{a} \neq \omega_{m}$) all four terms have explicit time dependence which, when averaged over many mechanical cycles, degrades the effective interaction strength. Conversely, on resonance ($\omega_{a} = \omega_{m}$) the interaction reduces to an atomic $\leftrightarrow$ mechanical state swap operation (since the `two-mode squeezing' terms of the lower line are far off-resonant)\footnote{Physically, the state swap terms correspond to in-phase operations which tend to correlate $\p_{a}\td$ and $\p_{m}\td$ with time-delayed versions of themselves ($\p_{a}\Br{t-\pi/\Omega}$ and $\p_{m}\Br{t-\pi/\Omega}$), whereas the two-mode squeezing terms correspond to in-quadrature operations which lead to correlations between the two oscillators.}. Sympathetic cooling leverages this fact by continually swapping the cold motional state of the atoms onto the mechanical device \cite{Hammerer2010}. We will therefore restrict ourselves to the special case of $\omega_{m} = \omega_{a} = \Omega$ in $\S$~\ref{Sec:Cooling}.

Note that the atomic $\leftrightarrow$ mechanical state swap which these dynamics perform is \textit{not} composed of consecutive swaps between the mechanical degrees of freedom and the light. A curious corollary is that the state transfer may be performed even in a regime where the optomechanical cooperativity is too low to allow an optical $\leftrightarrow$ mechanical swap. The requirements on experimental parameters (for the optomechanical device) are therefore significantly more relaxed than those of resolved-sideband cooling (\cf{} $\S$~\ref{Sec:AtomsAlone}).

Finally, we note that there are experimental advantages to incorporating a ring resonator into an experiment---despite the moderate increase in technical complexity---even though $g$ depends only on $\alpha_{L,R}$ (\ie{} on $\alpha_{d}/\sqrt{\kappa_{AC}}$). This is especially true if the drive strength $\alpha_{d}$ is limited (\eg{} by photodetector saturation or absorptive heating). For instance, keeping all other optical parameters constant (\ie{} fixed detuning, input power and transverse profile), our additional ring cavity yields a $g\propto\mathcal{\sqrt{\mathcal{F}_{AC}}}$ improvement of the coupling rate over that given by a free space trap \cite{Vogell2013} ($\mathcal{F}_{AC}$ is the finesse of the ring cavity, and we have imagined scaling $\omega_{m}\propto\sqrt{\mathcal{F}_{AC}}$ so as to maintain the $\omega_{a}=\omega_{m}$ resonance condition). Including a ring cavity may also permit one to use a larger transverse beam distribution (\eg{} in a bow-tie resonator), allowing $g$ to be boosted by trapping more atoms simultaneously. Alternatively, the cavity may be used to assist in suppressing heating of the atomic motion due to spontaneous photon scattering \cite{Gordon1980}, which occurs at a rate $\Gamma_{sc}$. To see this, note that $\Gamma_{sc}\propto\alpha_{R}\alpha_{L}/\Delta_{t}^{2} \propto \mathcal{F}_{AC}/\Delta_{t}^{2}$ (for fixed $\alpha_{d}$) \cite{Grimm2000} whilst $g\propto \Br{\mathcal{F}_{AC}/\Delta_{t}}^{3/4}$; we may therefore leave $g$ unchanged but suppress $\Gamma_{sc}$ by a factor of $1/\varepsilon$ by scaling both $\mathcal{F}_{AC}$ and $\Delta_{t}$ by $\varepsilon$ (see \ref{Sec:BackAction} for further discussion).

\subsection{\textbf{Coupling to reservoirs}} \label{Sec:Langevin}
Ultimately, the performance of our cooling scheme is limited by noise sources modelled by forming the Langevin equations \cite{GardinerZoller}
\begin{eqnarray}
Nm\sde{\hat{x}_{a}}{t} & = & -Nm\omega_{a}^{2}\x_{a} + K\x_{m} + \hat{F}_{CB} - \Gamma_{a}Nm\de{\x_{a}}{t}, \label{Eqn:MechACEff}\\
M\sde{\hat{x}_{m}}{t} & = & -M\omega_{m}^{2}\x_{m} +\hat{F}_{BA,m} + K\x_{a} + \hat{F}_{TH} - \Gamma_{m} M \de{\x_{m}}{t} + \hat{F}_{FB}. \label{Eqn:MechMTEff}
\end{eqnarray}
The atomic motion is damped into a cold bath (\eg{} by application of laser cooling \cite{Phillips1998}) at a rate $\Gamma_{a}$, which introduces fluctuations $\hat{F}_{CB}$. We will assume this reservoir to be at zero temperature. Mechanical losses are due to coupling to a hot bath at a rate $\Gamma_{m}$, with an associated forcing term $\hat{F}_{TH}$ consistent with a thermal occupancy $\bar{n}_{B,m}\approx k_{B}T_{B,m}/\hbar\omega_{m}\gg1$ (\cf{} \ref{Sec:Correlators}). Optical back-action on the mechanical oscillator is given by $\hat{F}_{BA,m}$, with the equivalent force on the atomic system vanishing within the realm of validity of our model (\cf{} \ref{Sec:Fields}). Finally, the effects of feedback are encapsulated by $\hat{F}_{FB}$.

It will be convenient to adopt a frequency domain description of the system for the purposes of treating the feedback circuit; for each time domain operator $\hat{f}\td$ there is a corresponding frequency domain operator $\tilde{f}\fd$ given by the Fourier transform
\begin{equation*}
\tilde{f}\fd = \intinf{t}\;\E^{\I\omega t}\hat{f}\td \Leftrightarrow \hat{f}\td = \frac{1}{2\pi}\intinf{\omega}\;\E^{-\I\omega t}\tilde{f}\fd.
\end{equation*}

Taking the transforms of Eqn~\eqref{Eqn:MechACEff} and Eqn~\eqref{Eqn:MechMTEff} and eliminating $\tilde{x}_{a}$ yields
\begin{equation}
\tilde{x}_{m}\fd = \frac{1}{\chi_{m}^{-1}\fd-K^{2}\chi_{a}\fd}\Sq{\tilde{F}_{TH}+\tilde{F}_{BA,m}+\tilde{F}_{FB} + K\chi_{a}\fd\tilde{F}_{CB}},
\label{Eqn:FourierXmNaive}
\end{equation}
in which the mechanical susceptibility is $\chi_{m}\fd = \Sq{M\Br{\omega_{m}^{2}-\omega^{2}-\I\omega\Gamma_{m}}}^{-1}$ and the atomic motion has the transfer function $\chi_{a}\fd = \Sq{Nm\Br{\omega_{a}^{2}-\omega^{2}-\I\omega\Gamma_{a}}}^{-1}$.

\subsection{\textbf{Modelling cold damping}} \label{Sec:Feedback} 

The fluctuation-dissipation theorem enforces our inability to cool an oscillator by increasing the rate at which it is damped into its thermal bath \cite{Pinard2000}; however, no such restrictions apply when coupled to a non-thermal environment. One method of engineering such an effective non-equilibrium reservoir is to use an external feedback circuit to apply a force $\hat{F}_{FB} \propto -\p_{m}$ to the oscillator, which increases its linewidth and introduces a (coloured) fluctuating force determined by noise on measurements of $\p_{m}$ \cite{Poggio2007}. The resulting mechanical steady-state is approximately thermal, with an effective temperature that may be less than that of the environment (\cf{} $\S$~\ref{Sec:ColdDamping}).

In optomechanical experiments $\p_{m}$ is typically not directly accessible; instead, one detects the phase quadrature of the output optical field (Eqn~\eqref{Eqn:OutputField}, with detection noise added as in \ref{Sec:Fields}), which carries information concerning $\x_{m}$, and feeds this signal through an electrical filter. Balanced homodyne is an appropriate detection method.

The most intuitive filter is a low-pass differentiator circuit\footnote{In practice one would also include a band-pass filter centred at $\omega_{m}$ so as to isolate the mechanical mode of interest.} with bandwidth $\Delta\omega_{FB} \gg \omega_{m}$ \cite{Mancini1998}; such a filter is optimal in the limit that $c_{m} \gg \bar{n}_{B,m}$ and $\eta = 1$, in that the controlled state asymptotically approaches the ground state as $c_{m}\rightarrow\infty$ (for appropriate choice of gain, \cf{} $\S$~\ref{Sec:ColdDamping}). The feedback force is
\begin{equation}
\tilde{F}_{FB} = \frac{\I\omega G \Gamma_{m}M}{1-\I\omega/\Delta\omega_{FB}} \tilde{x}_{m} + \tilde{F}_{SN}
\label{Eqn:FeedbackFunction}
\end{equation}
where the contribution due to optical shot noise is
\begin{equation*}
\tilde{F}_{SN} = \frac{-\I\omega G}{1-\I\omega/\Delta\omega_{FB}}\frac{M\Gamma_{m}\kappa_{MC}}{8 g_{m}\alpha_{d}}\Br{\delta\tilde{X}_{d}^{-}+\sqrt{\frac{1}{\eta}-1}\delta\tilde{Z}_{v}^{-}}.
\end{equation*}
The detection efficiency is $\eta \in \Sq{0,1}$ and $\delta\tilde{Z}_{v}^{-}$ is the uncorrelated vacuum noise coupled in by imperfect detection. We have normalised the filter function such that the overall feedback gain $G$ has a transparent physical interpretation (\cf{} Eqn~\eqref{Eqn:ModifiedSusceptibility}).

Inserting the feedback force into Eqn~\eqref{Eqn:FourierXmNaive} yields
\begin{equation}
\tilde{x}_{m}\fd = \chi_{m}^{\prime}\fd\Sq{\tilde{F}_{TH}+\tilde{F}_{BA,m}+\tilde{F}_{SN} + K\chi_{a}\fd\tilde{F}_{CB}}.
\label{Eqn:FourierXm}
\end{equation}

The effective mechanical susceptibility has been modified to
\begin{equation}
\chi_{m}^{\prime}\fd = \Sq{\chi_{m}^{-1}\fd - \frac{\I\omega}{1-\I\omega/\Delta\omega_{FB}} G M\Gamma_{m} -K^{2}\chi_{a}\fd}^{-1}.
\label{Eqn:ModifiedSusceptibility}
\end{equation}
In the limit $\omega_{m}\ll\Delta\omega_{FB}$ the second term describes a feedback-induced increase in the linewidth of the oscillator, with $G$ being the amount of broadening, whilst the third term is a modification due to the atomic centre of mass motion.

Feedback of the form described here may be implemented in a number of ways. For example, `electrostatic' actuation may be used to apply a force directly to the mechanical oscillator \cite{Hopkins2003,Poggio2007,Lee2010,Sridaran2011}, or the feedback force may be generated by imprinting an amplitude modulation onto a bright auxiliary optical field (which does not interact with the atomic system) \cite{Pinard2000,Kleckner2006,LIGO2009}. Importantly, we note that it is generally possible to arrange the feedback apparatus such that the quantum noise originating from the actuator (\eg{} the RF or optical field, respectively, for the examples given above) is negligible (\cf{} Eqn~\eqref{Eqn:FourierXm}); this approximation has been extensively employed in the optomechanical feedback literature (\eg{} \cite{Courty2001,Genes2008a,Doherty2012,Courty2000}).

\subsection{\textbf{Position power spectrum}} \label{Sec:CalculatingSxx}

The power spectrum of the position, $S\sb{x_{m}x_{m}}$, is found \via{} the Wiener-Khinchin theorem \cite{Aspelmeyer2013review} in the limit $\Delta\omega_{FB} \rightarrow \infty$.
\begin{eqnarray}
S\sb{x_{m}x_{m}}\Sq{\omega} & = & \frac{1}{2\pi}\intinf{\omega^{\prime}} \expect{\tilde{x}_{m}\fd \tilde{x}_{m}\left(\omega^{\prime}\right)} \nonumber \\
 & = & 2\hbar \modd{\chi^{\prime}_{m}}^{2} \left[ \Gamma_{m}M\omega_{m}\Br{\half + \bar{n}_{B,m}+c_{m}+\frac{\omega}{\omega_{m}}\frac{G}{4}+\frac{\omega^{2}}{\omega_{m}^{2}}\frac{G^{2}}{4^{2}\eta c_{m}}} \right. \nonumber\\
& \phantom{=} & \phantom{2\hbar \modd{\chi^{\prime}_{m}}^{2} \; \;} \left. + K^{2}\modd{\chi_{a}}^{2}\Gamma_{a}Nm\omega_{a}\Br{\half}\right]. \label{Eqn:PowerSpectrum}
\end{eqnarray}
The first two terms correspond to vacuum noise ($1/2$) and phonons entering the mechanics \via{} the thermal bath (with occupancy $\bar{n}_{B,m}$), whilst the third is the optomechanical cooperativity,
\begin{equation}
c_{m} = \frac{4\Br{g_{m}x_{zp,m}\beta}^{2}}{\Gamma_{m}\kappa_{MC}} = \frac{2\hbar}{M\omega_{m}\Gamma_{m}}\left(\frac{2g_{m}\alpha_{d}}{\kappa_{MC}}\right)^{2},
\label{Eqn:MechCooperativity}
\end{equation}
corresponding to the effective number of additional bath phonons introduced by optical back-action. As will be seen, $c_{m}$ controls the efficacy of feedback cooling \cite{Genes2008a} and strongly contributes to the sympathetic cooling performance.

The fourth term of Eqn~\eqref{Eqn:PowerSpectrum} arises from correlations between $\hat{F}_{BA,m}$ and $\hat{F}_{SN}$, and the fifth is solely due to the latter (\cf{} \ref{Sec:Correlators}).

Finally, noise entering from the zero-temperature bath is filtered by the atomic susceptibility and appears as the sixth term in the power spectrum. It is convenient to define an associated cooperativity by analogy with $c_{m}$:
\begin{equation}
c_{a} = \frac{2\hbar N}{m\omega_{a}\Gamma_{a}}\left(\frac{4kg_{a}\alpha_{d}}{\kappa_{AC}}\right)^{2} = \frac{Nm\omega_{a}^{3}}{2\hbar\Gamma_{a}}\Br{\frac{1}{4k\alpha_{d}}}^{2}.
\label{Eqn:AtomCooperativity}
\end{equation}
Although it does not appear directly in $S\sb{x_{m}x_{m}}$ (in the above form), $c_{a}$ is important in determining whether sympathetic cooling is capable of reaching the mechanical ground state.

Integrating over the power spectrum yields the steady-state variance of $\x_{m}$, \viz{}
\begin{equation}
\expect{\x_{m}^{2}} = \frac{1}{2\pi} \intinf{\omega}S\sb{x_{m}x_{m}}\Sq{\omega}, \label{Eqn:VarianceGeneral}
\end{equation}
which is the result of interest.

It is generally necessary to evaluate $\expect{\x_{m}^{2}}$ numerically; however, in $\S$~\ref{Sec:Cooling} we also explore several limits in which it is possible to give approximate analytical solutions of Eqn~\eqref{Eqn:VarianceGeneral}.

\section{Cooling Performance} \label{Sec:Cooling}
We are now in a position to calculate the mechanical oscillator's position variance as a function of the system parameters and the applied feedback gain. In the cases examined below the mechanical steady state is well approximated by a thermal distribution, as confirmed by numerical calculations of its covariance matrix. For this reason $\expect{\x_{m}^{2}}$ serves as an excellent proxy for temperature.

With this in mind, we will refer to the oscillator as `ground state cooled' if
\begin{equation}
\expect{\x_{m}^{2}} \leq 3x_{zp,m}^{2}.
\label{Eqn:GroundStateDefinition}
\end{equation}
This corresponds to the requirement that the oscillator contain at most one phonon on average ($\expect{\bcop{a}_{m}\baop{a}_{m}} \leq 1$) \footnote{We note that this is the same cut-off as implicitly employed in previous works \eg{} \cite{Hammerer2010,Vogell2013}}.

\subsection{\textbf{Feedback Cooling}} \label{Sec:ColdDamping}

Let us first consider the case in which there are no atoms in the trap. We include a brief summary of results (\cf{} sources presented in $\S$~\ref{Sec:Introduction}) here for the purposes of comparison with sympathetic cooling.

It is straightforward to extremise Eqn~\eqref{Eqn:VarianceGeneral} with $c_{a}=0$ (using the relations given in \ref{Sec:Maths}), confirming the presence of a global minimum variance of
\begin{equation}
\frac{\expect{\x_{m}^{2}}_{opt}}{x_{zp,m}^{2}} = \frac{G_{opt}^{\Br{0}}}{4 \eta c_{m}}
\label{Eqn:VarianceFeedback}
\end{equation}
which is achieved with the feedback gain $G_{opt}^{\Br{0}} = \sqrt{1+\SNR}-1$, where the signal-to-noise ratio \cite{Lee2010} is given by $\SNR = 16\eta c_{m}\Br{\bar{n}_{B,m}+c_{m}+1/2}$.

In the experimentally-relevant regime of $\left\{\SNR,\bar{n}_{B,m}\right\} \gg 1$ it is straightforward to show that ground state cooling may be realised if
\begin{equation}
\bar{n}_{B,m} \lesssim \Br{9\eta-1}c_{m} \leq 8 c_{m}.
\label{Eqn:FeedbackGroundstateCriterion}
\end{equation}
Thus we see that feedback cooling to the ground state is possible when the mechanical noise spectrum is dominated by radiation pressure fluctuations.

\subsection{\textbf{Sympathetic cooling}} \label{Sec:AtomsAlone}

We now consider the capacity of sympathetic cooling alone: our analysis complements and extends those performed by \cite{Hammerer2010} and \cite{Vogell2013} by explicitly determining the temperature achievable in the regime of hybridised mechanical modes.

Analytical treatments are tractable in both the atomic adiabatic (weak coupling) regime, wherein the atoms are damped heavily compared to the rate of phonon transfer between the oscillators, and the strong coupling (hybridised) regime, in which the coherent interaction (Eqn~\eqref{Eqn:EffectiveHamiltonian}) is dominant. The primary challenge in either case is evaluating $\intinf{\omega} \modd{\chi_{m}^{\prime}}^{2}$ and $\intinf{\omega} \modd{\chi_{m}^{\prime}}^{2}\modd{\chi_{a}}^{2}$. These control, respectively, the response to noise acting directly on the mechanics and indirectly \via{} the atoms.

In the weak coupling regime ($\Gamma_{a} \gg g$ \& $\Gamma_{a} > \Gamma_{m}$) the atomic centre of mass adiabatically follows the motion of the mechanical oscillator, allowing us to expand the atomic susceptibility near the resonance frequency as
\begin{equation*}
\chi_{a}\Br{\omega \approx \Omega} \approx \frac{\I}{Nm\omega\Gamma_{a}} \approx \frac{\I}{Nm\Gamma_{a}\Omega}\Br{2-\frac{\omega}{\Omega}}.
\end{equation*}
Neglecting the small frequency-independent imaginary term which arises leaves us with the approximate modified mechanical susceptibility
\begin{equation}
\chi^{\prime}_{m}\left(\omega\right) \approx \frac{1}{M\Sq{\Omega^{2}-\omega^{2}-\I\Gamma_{m}\omega\left(1+c\right)}}
\label{Eqn:EffectiveChi}
\end{equation}
which describes a simple harmonic oscillator with an enhanced linewidth of $\Gamma_{m}\left(1+c\right)$. Fittingly, the broadening is characterised by the atom--mechanical cooperativity
\begin{equation}
c = \frac{4 g^{2}}{\Gamma_{a}\Gamma_{m}} = 4^{2}c_{a}c_{m},
\label{Eqn:Cooperativity}
\end{equation}

With these approximations $\modd{\chi_{m}^{\prime}}^{2}$ may be readily integrated (\cf{} \ref{Sec:Maths}).

The remaining term, proportional to $\intinf{\omega} \modd{\chi_{m}^{\prime}}^{2}\modd{\chi_{a}}^{2}$, is not evaluated directly. Instead, as shown in \ref{Sec:ApproxAnalytical}, we approximate the integrand around the mechanical resonance frequency as a Lorentzian peak; since the majority of the spectral variance is contained within a relatively narrow bandwidth the true integral is faithfully reproduced.

Using these two approximations we find
\begin{equation}
\frac{\expect{\x_{m}^{2}}}{x_{zp,m}^{2}} = \frac{2}{1+c}\Sq{ \bar{n}_{B,m}+c_{m}+\half+\bar{n}_{a,eff} },
\label{Eqn:VarianceWeak}
\end{equation}
where coupling to the atoms has introduced an effective number of phonons
\begin{equation*}
\bar{n}_{a,eff} = \frac{c/2}{1+\frac{\Gamma_{m}}{\Gamma_{a}}\Br{1+c}}.
\end{equation*}

In the opposite limit, strong coupling ($g \gg \max\Cu{\Gamma_{a},\Gamma_{m}}$), excitations are hybridised across the two local modes ($\x_{a}$ and $\x_{m}$), giving rise to a symmetric and an antisymmetric normal mode. In this case, as detailed in \ref{Sec:ApproxAnalytical}, $\modd{\chi_{m}^{\prime}}^{2}$ and $\modd{\chi_{m}^{\prime}}^{2}\modd{\chi_{a}}^{2}$ share a similar twin-peaked structure. In both instances we exploit the rapid decay of the spectral variance away from these peaks, much as above, to integrate Eqn~\eqref{Eqn:PowerSpectrum}, yielding
\begin{equation}
\frac{\expect{\x_{m}^{2}}}{x_{zp,m}^{2}} = \half\Sq{\frac{\Gamma_{m}}{\Gamma_{N}}\Br{\bar{n}_{B,m} + c_{m}} + 1}{\Br{1-\frac{g^{2}}{\Omega^{2}}}^{-2}\Br{1-\frac{4g^{2}}{\Omega^{2}}}}^{-1}.
\label{Eqn:VarianceStrong}
\end{equation}
Here $\Gamma_{N} = \half\Br{\Gamma_{a}+\Gamma_{m}}$ is the linewidth of the normal modes of the system.

\begin{figure}[bt]
\begin{center}
\def\svgwidth{1\columnwidth}
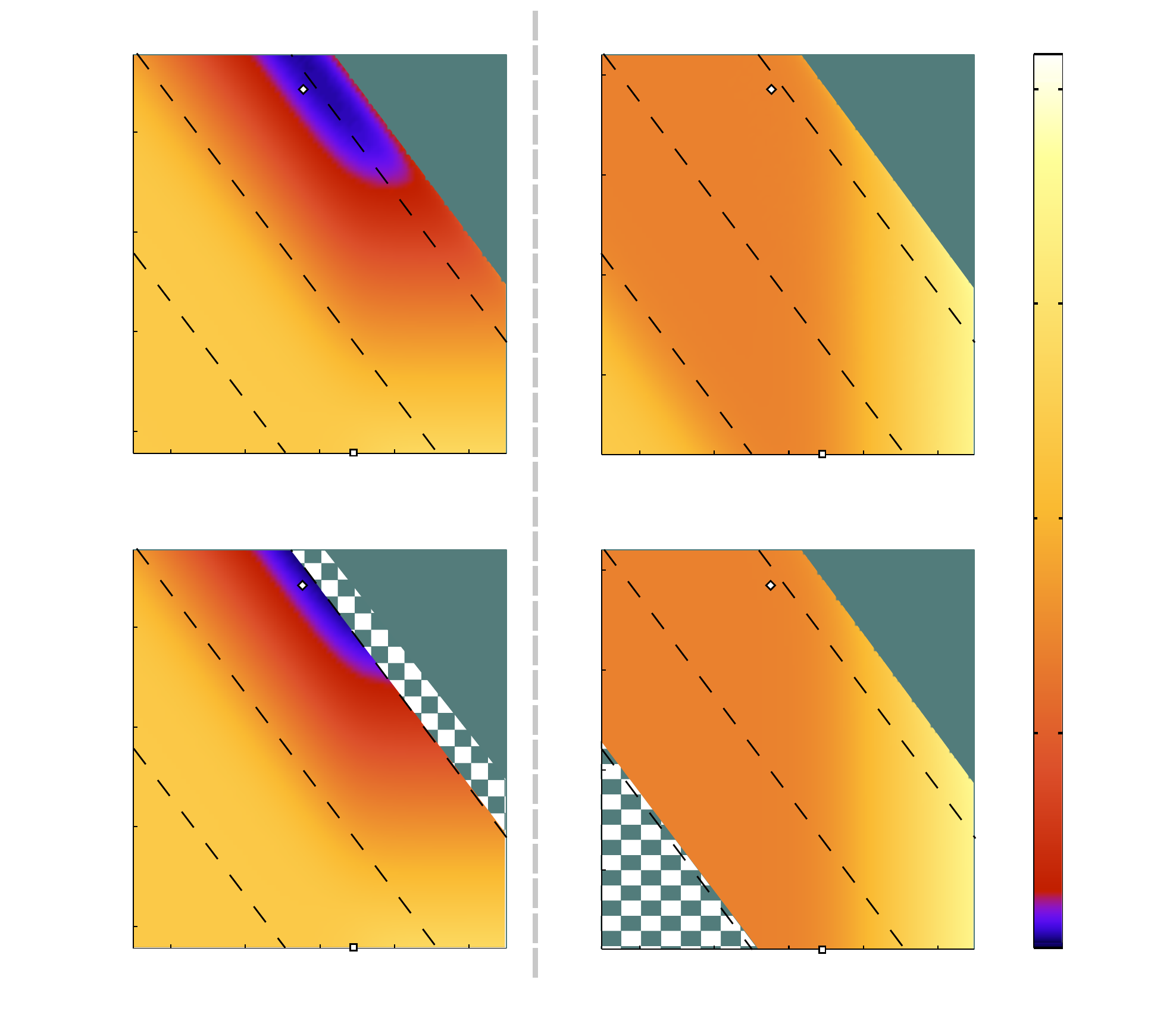
\caption{\label{Fig:Sympathetic} The temperature of a micromechanical device sympathetically cooled by coupling to a cold, trapped atomic gas. Purple indicates cooling to near the motional ground state. Contours of equal $g/\Omega$ are indicated by dashed lines, and the unstable $g > \Omega/2$ region is coloured teal (\cf{} Eqn~\eqref{Eqn:Instability}). The horizontal axis has been normalised (\cf{} Eqn~\eqref{Eqn:FeedbackGroundstateCriterion}). Numerical results are presented in the upper row and the corresponding analytical predictions in the lower. See the text for discussion of the input parameters used. The squares and diamonds indicate feasible experimental parameters (\cf{} Table~\ref{Tab:Params}). \newline
A) $\Gamma_{a} = \Omega$. Sympathetic cooling alone is capable of reaching near-ground-state temperatures in the weak coupling regime. Note that the cold `tongue' protrudes well into the $8c_{m} > \bar{n}_{B,m}$ region in which feedback cooling is also capable of approaching the ground state.\newline
B) $\Gamma_{a} = \Omega$. Eqn~\eqref{Eqn:VarianceWeak} accurately predicts the true steady-state variance in the intended parameter space. In the checked region the system is hybridised \ie{} strongly coupled.\newline
C) $\Gamma_{a} = 1.4\times10^{-5}\;\Omega  = 10^{2}\;\Gamma_{m}$. As discussed in the text, near-ground-state cooling is not possible if the atoms are incapable of dissipating energy faster than it enters from the mechanical system.\newline
D) $\Gamma_{a} = 1.4\times10^{-5}\;\Omega  = 10^{2}\;\Gamma_{m}$. A second analytical expression (Eqn~\eqref{Eqn:VarianceStrong}) is accurate in the strong coupling regime. The equation is invalid in the checked region, which is in fact weakly coupled ($g < \Gamma_{a}$).}
\end{center}
\end{figure}

Our analytical expressions (Eqn~\eqref{Eqn:VarianceWeak} and Eqn~\eqref{Eqn:VarianceStrong}) are compared to numerical results in Fig.~\ref{Fig:Sympathetic}. The parameters $\Omega$ and $\Gamma_{m}$ have been chosen to be representative of a SiN nanostring \cite{Schmid2011,Brawley2012} ($\Omega/2\pi = 220$ kHz, $\Gamma_{m} = 195$ mHz) held in a dilution refrigerator such that $\bar{n}_{B,m} = 2.8\times10^{4}$ (temperature $T_{B,m} = 300$ mK, \cf{} \cite{Thompson2008}). Optical dipole traps are readily capable of achieving vibrational frequencies on the order of $\Omega$ \cite{Camerer2011, Vetsch2012}. The remaining independent\footnote{Of course, the cooperativities \textit{do} depend on the resonance frequency and decay rates: however, it is possible to vary them independently by suitable adjustment of $g_{a}$, $N$, $g_{m}$, \etc{}} parameters appearing in the power spectrum are $c_{a}$, $c_{m}$ and $\Gamma_{a}$ (with $G$ and $\eta$ being important when measurement feedback is included). These parameters are discussed further in Table~\ref{Tab:Params} and $\S$~\ref{Sec:CombinedCooling}.

We may now ask whether sympathetic cooling is capable of producing near-ground-state-cooled mechanical oscillators.

Substituting Eqn~\eqref{Eqn:VarianceWeak} into the ground state criterion (Eqn~\eqref{Eqn:GroundStateDefinition}) and using the fact that $\bar{n}_{B,m} \gg 1$ yields
\begin{equation}
\bar{n}_{B,m} < \Br{2+3\frac{\Gamma_{m}}{\Gamma_{a}}}\bar{n}_{a,eff}-c_{m}
\label{Eqn:AtomicGroundstateCriterion}
\end{equation}
as a sufficient condition for ground state cooling in the adiabatic limit (we have used $c \Gamma_{m}/\Gamma_{a} \ll 1$, which holds in this case). Much as the condition $\bar{n}_{B,m} < 8c_{m}$ indicates that the mechanical spectrum must be dominated by radiation pressure fluctuations in order to ground state cool using feedback, Eqn~\eqref{Eqn:AtomicGroundstateCriterion} shows that \textit{atomic} contributions must dominate in order to sympathetically cool to the ground state.

It is straightforward to show that the region in which Eqn~\eqref{Eqn:AtomicGroundstateCriterion} is satisfied is a portion of the parameter space where $c > \bar{n}_{B,m}$ and $c_{a} > 1/24$. The condition $c > \bar{n}_{B,m}$ is of great importance, as it also dictates whether near-ground-state cooling may be achieved in the strong coupling regime.

Inserting the variance in the hybridised regime (Eqn~\eqref{Eqn:VarianceStrong}) into the ground state criterion gives an inequality which may only be satisfied if $c = \bar{n}_{B,m}$ is reached in the \textit{weak} coupling limit; that is, if near-ground-state cooling is possible in the adiabatic limit then it is also possible in the case of strong coupling\footnote{We imagine tuning between the two regimes by varying $c_{a}$ and $c_{m}$, keeping $\Gamma_{a}, \Gamma_{m}$ and $\Omega$ fixed.}. The converse statement is also true; if near-ground-state cooling is \textit{not} possible in the weakly-coupled limit then it is also \textit{not} possible in the hybridised limit (\ie{} no near-ground cooling if $c=\bar{n}_{B,m}$ requires strong coupling).

The physical interpretation of these statements is that the atomic motion must be damped into the zero-temperature bath faster than phonons enter from the hot reservoir, which is entirely consistent with our na\"{i}ve expectations. Effective steady-state cooling is therefore more difficult to realise in the strongly-coupled case simply because $\Gamma_{a}$ is bounded above by $\Omega/2$ (required for stability, \cf{} Eqn~\eqref{Eqn:Instability}).

In summary, we have shown that there exists two (slightly overlapping) parameter regimes, summarised in Table~\ref{Tab:Regimes}, in which the oscillator is prepared near the quantum ground state.

\begin{table}
\centering
\begin{tabular}{l | l}
Near-ground cooling & Condition \\
\hline
Sympathetic & $\bar{n}_{B,m} < c$ \& $1/24 < c_{a}$ \& $\Gamma_{m}\bar{n}_{B,m} \ll \Gamma_{a}$\\
Feedback & $\bar{n}_{B,m} < \Br{9\eta-1}c_{m}$\\
Neither & $\max\Cu{\Br{9\eta-1}c_{m},c}<\bar{n}_{B,m}$ or $c_{a} < 1/24$\\
\hfill & or $\Gamma_{a} \lesssim \Gamma_{m}\bar{n}_{B,m}$ \\
\end{tabular}
\caption{\label{Tab:Regimes} A summary of the relevant parameter regimes for sympathetic and feedback cooling to near the ground state. In the case that both sympathetic cooling and cold damping are capable of approaching the ground state there exists an overlap region if the feedback efficiency satisfies $1 \geq \eta > \frac{1}{9}\Br{1+\frac{2}{3}\Gamma_{a}\Gamma_{m}\bar{n}_{B,m}\Omega^{-2}}$.}
\end{table}

\FloatBarrier

\subsection{\textbf{Combined cooling}} \label{Sec:CombinedCooling}

We now turn our attention to the performance of combined sympathetic and feedback cooling. Both mechanisms act to suppress the leakage of information into the environment; cold damping achieves this explicitly by measurement and feedback, whilst atomic cooling achieves the same effect by diverting a portion of the leakage into the atoms, and thence back to the mechanical system. Our main computational task is therefore to determine the optimum feedback gain to apply for a fixed sympathetic cooling capacity, and to then calculate the new---hopefully reduced---position variance.

The former is typically pushed away from its $c_{a}=0$ value, $G_{opt}^{\Br{0}}$, to a new optimum, $G_{opt}$. To calculate this gain we analytically differentiate $S\sb{x_{m}x_{m}}$ with respect to $G$, numerically integrate over $\omega$ to find $\partial \expect{x_{m}^{2}} / \partial G$, and apply a numerical root-finding algorithm to determine $G_{opt}$.

\begin{figure}[bt]
\begin{center}
\def\svgwidth{1\columnwidth}
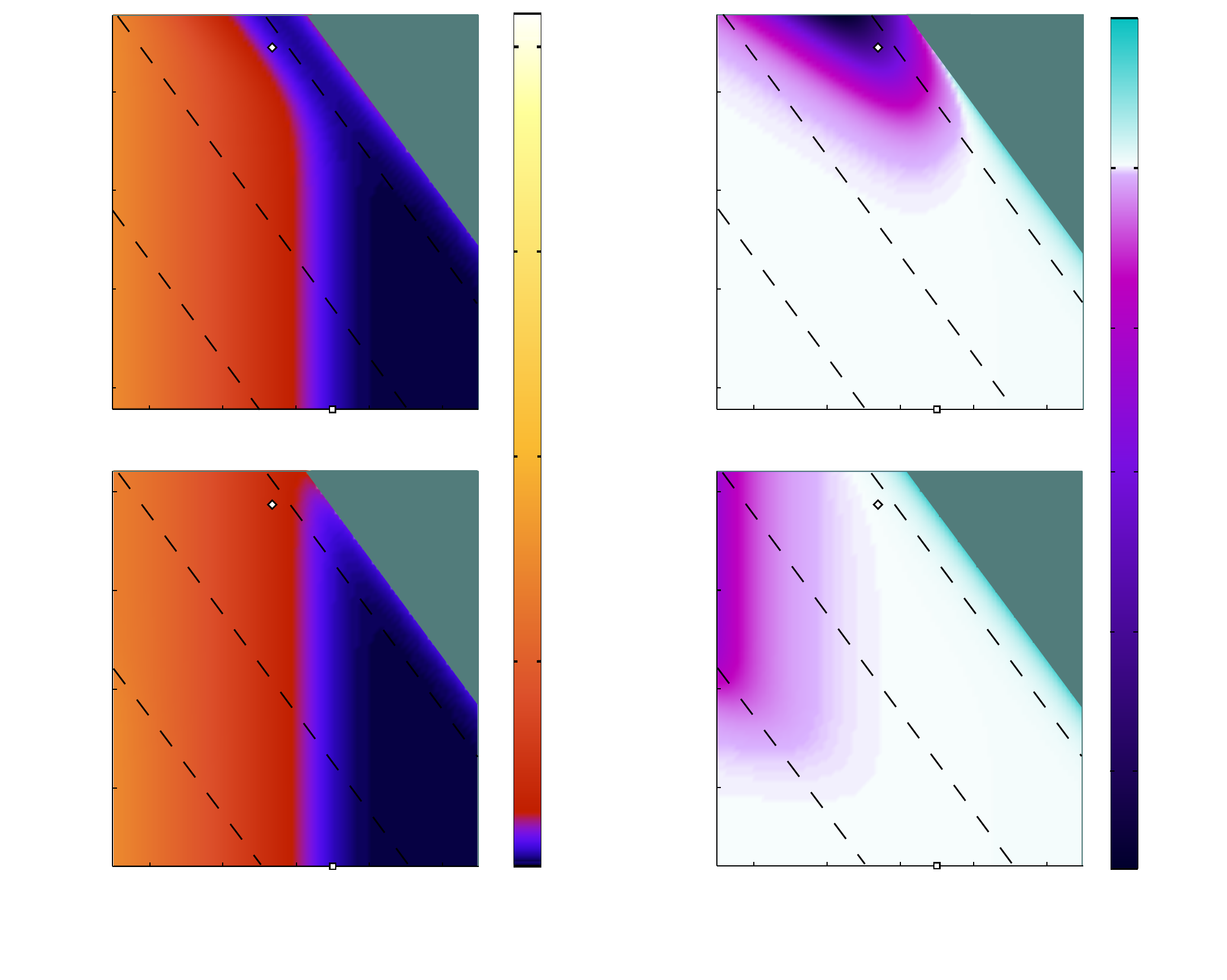
\caption{\label{Fig:CombinedCooling} The steady-state behaviour of a mechanical device subject to combined sympathetic and feedback cooling ($\eta = 1$). The left column shows numerical calculations of the optimum position variance, and the right displays the corresponding optimised feedback gain ($G_{opt}$) relative to its $c_{a} = 0$ value ($G_{opt}^{\Br{0}}$). Note that the colour scale on A) and B) matches that of Fig.~\ref{Fig:Sympathetic}, as do the annotations. These results have been truncated near to the instability region because numerical calculations of $\partial\expect{\x_{m}^{2}}/\partial G$ become unreliable in this limit. Squares and diamonds indicate achievable experimental parameters, as in Table~\ref{Tab:Params}. \newline
A) $\Gamma_{a} = \Omega$. Feedback does not significantly alter the optimum temperature in the region where the atoms alone are capable of ground state cooling; however, it is of significant use outside of this area.\newline
B) $\Gamma_{a} = 1.4\times10^{-5} \;\Omega  = 10^{2}\;\Gamma_{m}$. If feedback is introduced in the regime $\Gamma_{m} \sim \Gamma_{a}$ it is possible to reach much smaller variances than with atoms alone.\newline
C) $\Gamma_{a} = \Omega$. The optimum gain $G_{opt}$ is not appreciably altered from its atom-free value $G_{opt}^{\Br{0}}$ across a wide range of parameters. The dark region in the upper left of the plot indicates that the atoms are dominating cooling in this regime.\newline
D) $\Gamma_{a} = 1.4\times10^{-5}\;\Omega  = 10^{2}\;\Gamma_{m}$. Deviations from $G_{opt}^{\Br{0}}$ occur for low $c_{m}$ because the feedback circuit becomes unable to track the mechanical position with sufficient accuracy. In this regime sympathetic cooling plays an increasingly significant role.}
\end{center}
\end{figure}

\begin{figure}[bt]
\begin{center}
\def\svgwidth{1\columnwidth}
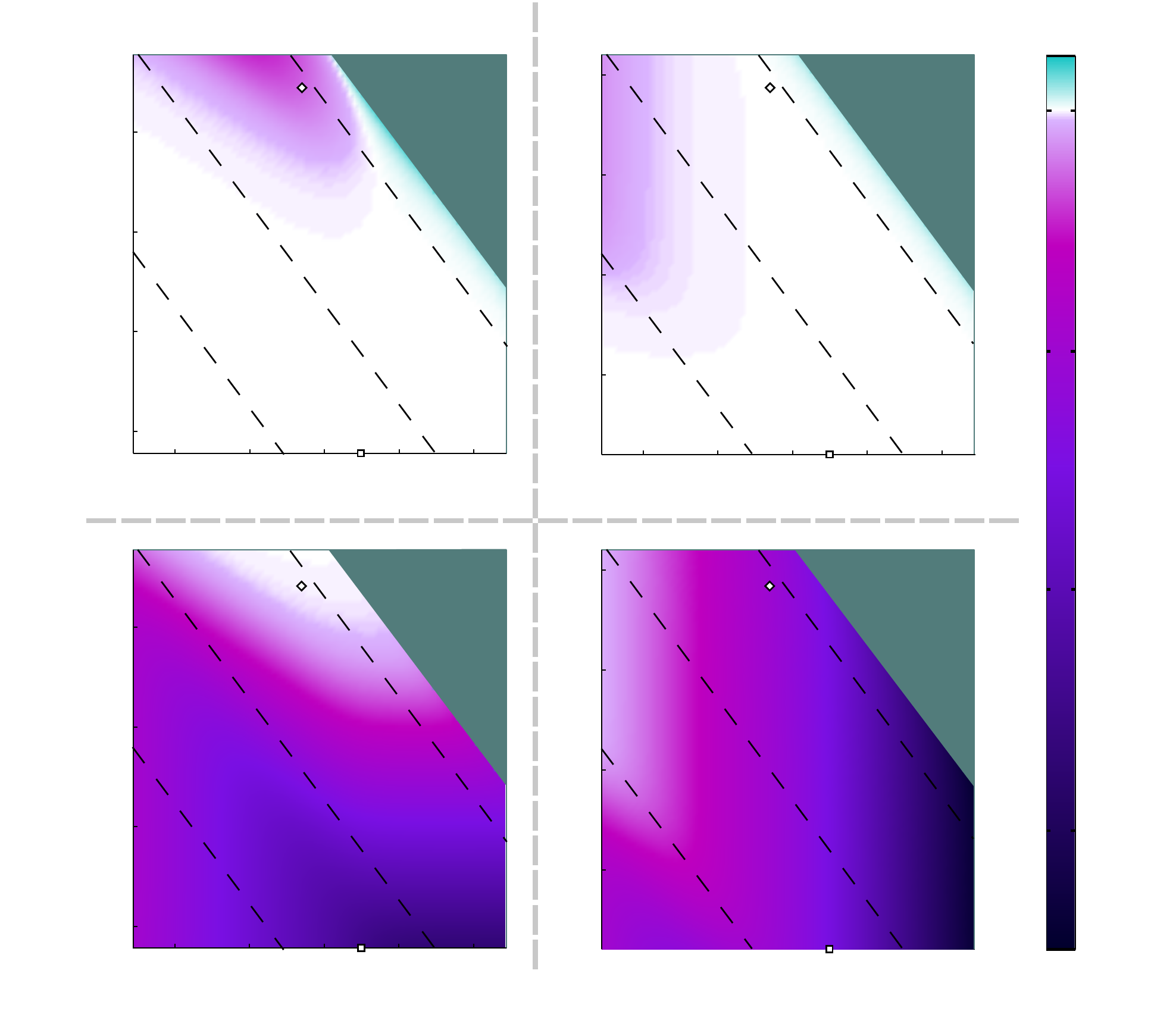
\caption{\label{Fig:Comparison} A comparison of combined feedback and sympathetic cooling with each mechanism employed individually. The left column shows results in the weakly-coupled regime, with strong on the right. The upper row compares the variance with combined cooling to that with optimised feedback in the absence of atoms (\ie{} $\expect{\x_{m}^{2}\Br{c,G_{opt}}}/\expect{\x_{m}^{2}\Br{0,G_{opt}^{\Br{0}}}}$, in dB), whilst the lower column shows the performance of combined cooling with respect to sympathetic cooling only (\ie{} $\expect{\x_{m}^{2}\Br{c,G_{opt}}}/\expect{\x_{m}^{2}\Br{c,0}}$, in dB). Feasible experimental parameters (\cf Table~\ref{Tab:Params}) are plotted as squares and diamonds. \newline
A) $\Gamma_{a} = \Omega$. Addition of atoms in the weakly-coupled regime does allow one to reach temperatures below that achieved by feedback alone, with the greatest impact being for parameter combinations where feedback cooling is incapable of approaching the ground state.\newline
B) $\Gamma_{a} = \Omega$. Introducing feedback cooling dramatically reduces the variance everywhere outside of the near-ground-state-cooled area.\newline
C) $\Gamma_{a} = 1.4\times10^{-5}\;\Omega = 10^{2}\;\Gamma_{m}$. The cooling is almost completely dominated by the measurement-based feedback in the strong coupling regime. \newline
D) $\Gamma_{a} = 1.4\times10^{-5}\;\Omega = 10^{2}\;\Gamma_{m}$. The sympathetic damping mechanism contributes most strongly to the combined cooling power of the hybrid system in the low-$c_{m}$ regime.}
\end{center}
\end{figure}

The results shown in Fig.~\ref{Fig:CombinedCooling} and Fig.~\ref{Fig:Comparison} are calculated in the case of perfect feedback \ie{} $\eta = 1$: we note that the results with imperfect feedback efficiency are qualitatively the same for $\eta \gtrsim 50\%$ (one essentially need only renormalise the lower axis appropriately), but differ substantially for $\eta \lesssim 15\%$.

These data make it quite clear that, generally speaking, including measurement-based feedback alongside sympathetic cooling significantly decreases the oscillator's temperature. The notable exception to this behaviour is in the parameter space where atoms alone are capable of achieving near-ground-state temperatures ($8 c_{m} < \bar{n}_{B,m} < c$ \& $\Gamma_{m}\bar{n}_{B,m} \ll \Gamma_{a}$), in which the addition of feedback yields little improvement ($\sim 0.01$ dB improvement). Furthermore, the addition of atoms to the system has negligible impact if feedback cooling to the ground state is possible.

Exemplary experimental parameters are given in Table~\ref{Tab:Params}; these yield the points denoted by $\square$ and $\diamond$ in Figures \ref{Fig:Sympathetic}, \ref{Fig:CombinedCooling} and \ref{Fig:Comparison}.

Ideally, our example mechanical system which operates in the mechanical back-action--dominated regime ($8c_{m}>\bar{n}_{B,m}$, left panel of Table~\ref{Tab:Params}) will reach a final variance of $1.41\,x_{zp,m}^{2}$ with feedback alone: its thermal variance is $\sim 5\times 10^{4}\,x_{zp,m}^{2}$.

Our suggested hybrid system (right panel of Table~\ref{Tab:Params}: note that $g_{m}$ has been adjusted such that $8c_{m}<\bar{n}_{B,m}$ \ie{} cold damping to the ground state is not possible) achieves variances of $1.35\,x_{zp,m}^{2}$ and $1.33\,x_{zp,m}^{2}$ in the weakly-coupled regime---with and without feedback, respectively---whilst in the case of strong coupling the (sympathetically-cooled) variance of $603\,x_{zp,m}^{2}$ may be reduced to a mere $4.79\,x_{zp,m}^{2}$ by switching on feedback. This is essentially equal to the feedback-only steady-state variance in this regime.

It is encouraging that all of these parameters are within reach of state-of-the-art optomechanical and atomic systems; sympathetic cooling (and/or cold damping) to the mechanical ground state is technically feasible. The experimental challenges lay in combining these heretofore disparate elements and eliminating technical noise (and system-specific noise sources, such as absorptive heating) which we have not considered here.

\begin{table}
\centering
\begin{tabular}{l| l}
Mechanical alone \small{$\square$} \cf{} \cite{Brawley2012}
& Mechanical and atomic $\diamond$ \\
\hline
	\begin{tabular}{c l}
		$\omega_{m}/2\pi$ & $220$~kHz \\
		$\Gamma_{m}/2\pi$ & $31$~mHz \\
		$M$ & $1.4$~ng \\
		$\kappa_{MC}$ & $20\:\omega_{m}$ \\
		$\alpha_{d}$ & $6.58\times 10^{6}$~$\sqrt{\mathrm{Hz}}$ \\
		$g_m$ & $9.85$~MHz/nm\\
		$T_{B,m}$ & $300$~mK \cite{Blencowe2004} \\
		$8c_{m}/\bar{n}_{B,m}$ & $9.03$~dB \\
	\end{tabular}
&
	\begin{tabular}{c l | c l}
		$\omega_{m}/2\pi$ & $220$~kHz & $\omega_{a}$ & $\omega_{m}$ \\
		$\Gamma_{m}/2\pi$ & $31$~mHz & $\Gamma_{a}$ & $\omega_{m}~\Br{10^{2}\:\Gamma_{m}}$ \\
		$M$ & $1.4$~ng & $m$ & $1.44\times 10^{-25}$~kg \\
		$\kappa_{MC}$ & $20\:\omega_{m}$ & $\kappa_{AC}$ & $20\:\omega_{m}$ \\
		$\alpha_{d}$ & $6.58\times 10^{6}$~$\sqrt{\mathrm{Hz}}$ & $N$ & $3.1\times 10^{8}$ \cite{Kerman2000} \\
		$g_m$ & $3.19$~MHz/nm & $\Delta_{t}/2\pi$ & $-1$~GHz \cite{Vogell2013} \\
		$T_{B,m}$ & $300$~mK \cite{Blencowe2004} & $\mathcal{V}$ & $2.8\times 10^{-8}$~$\mathrm{m}^{3}$ \\
		$8c_{m}/\bar{n}_{B,m}$ & $-4.38$~dB & $c_{a}$ & $9.54~\Br{58.1}$~dB \\
	\end{tabular} \\
\end{tabular}
\caption{\label{Tab:Params} Feasible experimental parameters which permit preparation of a mechanical oscillator near its ground state by using sympathetic or feedback cooling.\newline
The mechanical ($\square$) specifications are drawn from the literature concerning evanescently-coupled, high-tension silicon nitride nanostrings (as discussed in $\S$~\ref{Sec:AtomsAlone}); we have assumed that the damping rate is independent of temperature.\newline
The atomic cavity ($\diamond$) parameters have been chosen to be comparable to those used to construct optical parametric oscillators (\eg{} cavity length $\sim 30$~cm, linewidth $\kappa_{AC}/2\pi = 4.4$~MHz). The transverse beam area is drawn from \cite{Camerer2011} and the detuning estimated according to \cite{Vogell2013}. We have used the transition wavelength and dipole moment of the \isotope{Rb}{87} D2 line \cf{} \cite{Vogell2013}. Values in parentheses are valid in the small $\Gamma_{a}$ limit. It is possible to prepare large numbers of atoms in their motional ground state (\eg{} using Raman cooling in a three-dimensional optical lattice) \cite{Kerman2000}. \newline
See $\S$~\ref{Sec:CombinedCooling} for discussion of the final temperatures achieved using these specifications.}
\end{table}

\FloatBarrier

\section{Conclusion} \label{Sec:Conclusion}

We have modelled steady-state cooling of a low-frequency mechanical oscillator using the combined effects of optical coupling to a remote atomic ensemble and measurement-based feedback. Combining these two methods is beneficial in all circumstances, although there exists distinct regions of parameter space (\cf{} Table \ref{Tab:Regimes}) in which one technique or the other dominates the cooling. We have also demonstrated that an optically-mediated state swap between the two mechanical degrees of freedom (\ie{} sympathetic cooling to the ground state) may be performed even in the case that $\bar{n}_{B,m} < c_{m}$, in which a complete mechanical $\leftrightarrow$ optical swap is forbidden. Both sympathetic and feedback cooling to the ground state are feasible with current experimental parameters.

\section{Acknowledgements}
The authors extend their thanks to George A. Brawley and Glen I. Harris for technical assistance, and Klemens Hammerer and Philipp Treutlein for useful discussions.  This work was supported by the Australian Research Council (EQuS, CE$110001013$). Lars S. Madsen further acknowledges support from the Carlsberg Foundation.

\pagebreak
\section{References}

\providecommand{\newblock}{}

\pagebreak
\appendix
\section[\hspace{3cm} Treatment of adiabatic field evolution]{Treatment of adiabatic field evolution} \label{Sec:Fields}

As explained in $\S~$\ref{Sec:Theory}, we assume that the system is operating in a regime where the field fluctuation operators rapidly reach quasistatic forms relative to the mechanical period(s). Furthermore, we suppose that both optical cavities are strongly overcoupled ($\kappa_{AC} \gg \left\{\Gamma_{AC},-2Nkg_{a}x_{zp,a}\right\}$ and $\kappa_{MC} \gg \left\{\Gamma_{MC},g_{m}x_{zp,m}\right\}$) such that perturbations of the steady-state field amplitudes away from their interaction-free values are negligibly small. In this limit we have $\alpha_{L} = \alpha_{R} = 2\alpha_{d}/\sqrt{\kappa_{AC}}$ and $\beta = 2\alpha_{d}/\sqrt{\kappa_{MC}}$.

\subsection[\hspace{3cm} Adiabatic limit of the optical Langevin equations]{\textbf{Adiabatic limit of the optical Langevin equations}}

We describe the fields in the optical cavities with the Langevin equation \cite{WallsMilburn2008}
\begin{equation}
\de{\delta \baop{a}}{t} = \frac{1}{\I\hbar}\comm{\delta \baop{a}}{\hat{H}} + \sqrt{\kappa}\delta\baop{a}_{in}- \half\kappa\delta \baop{a},
\label{Eqn:OpticalLangevin}
\end{equation}
in which the coupling rate\footnote{It may be shown that including extraneous losses (\textit{e.g.} scattering into unguided modes and absorption loss) at a rate $\Gamma$ leads to an amount of additional noise which scales as $\Gamma/\kappa$; we therefore neglect these contributions in the regime $\kappa\gg\Gamma$.} between the resonator and drive mode $\delta \baop{a}_{in}$ is $\kappa$, and $\hat{H}$ is either Eqn~\eqref{Eqn:AtomLightLin} or Eqn~\eqref{Eqn:OptoMechLin} as appropriate. $\delta \baop{a}_{in}$ represents the fluctuations of the multi-mode input field.

Specifically, the fluctuations of the drive mode obey the relationships \cite{Milburn2011}
\begin{eqnarray*}
\expect{\delta\baop{a}_{d}} & = & 0, \\
\expect{\delta\bcop{a}_{d}\td\delta\baop{a}_{d}\left(t^{\prime}\right)} & = & 0, \\
\expect{\delta\baop{a}_{d}\td\delta\bcop{a}_{d}\left(t^{\prime}\right)} & = & \delta\left(t-t^{\prime}\right). \\
\end{eqnarray*}
These translate into correlations between sidebands in the frequency domain (the carrier having been translated to $\omega = 0$).
\begin{eqnarray*}
\expect{\delta \tilde{X}_{j}^{\pm}\fd \delta \tilde{X}_{k}^{\pm}\left(\omega^{\prime}\right)} & = & 2\pi \delta_{j,k}\delta\left(\omega+\omega^{\prime}\right),\\
\expect{\delta \tilde{X}_{j}^{\pm}\fd \delta \tilde{X}_{k}^{\mp}\left(\omega^{\prime}\right)} & = & \pm \I 2\pi \delta_{j,k}\delta\left(\omega+\omega^{\prime}\right). \\
\end{eqnarray*}

Transfer of information between the two optical cavities is treated using the input-output formalism. The appropriate input-output relations are (\cf{}~Fig.~\ref{Fig:ModelGeometry}) \cite{Milburn2011}
\begin{eqnarray*}
\delta \baop{a}_{in,R} & = & \delta \baop{a}_{d}, \\
\delta \baop{b}_{in} & = & \sqrt{\kappa_{AC}}\delta \baop{a}_{R} - \delta \baop{a}_{d}, \\
\delta \baop{a}_{in,L} & = &  \sqrt{\kappa_{MC}}\delta \baop{b} - \delta \baop{b}_{in}. \\
\end{eqnarray*}
These expressions are valid in the case that the time taken for light to propagate between the two cavities is small compared to the mechanical period(s).

Application of Eqn~\eqref{Eqn:OpticalLangevin} in the adiabatic limit yields 
{\begin{eqnarray*}
	\delta \tilde{X}_{R}^{+} & = & \frac{2}{\sqrt{\kappa_{AC}}} \left(\delta \tilde{X}^{+}_{d}+\frac{4Nkg_{a}\alpha_{L}}{\sqrt{\kappa_{AC}}}\tilde{x}_{a}\right), \\
	\delta \tilde{X}_{R}^{-} & = & \frac{2}{\sqrt{\kappa_{AC}}} \Br{\delta\tilde{X}_{d}^{-}}, \\
	\delta \tilde{Y}^{+} & = & \frac{2}{\sqrt{\kappa_{MC}}} \Br{\delta\tilde{X}_{d}^{+}+\frac{8Nkg_{a}\alpha_{L}}{\sqrt{\kappa_{AC}}}\tilde{x}_{a}}, \\
	\delta \tilde{Y}^{-} & = & \frac{2}{\sqrt{\kappa_{MC}}} \Br{\delta\tilde{X}_{d}^{-}-\frac{2g_{m}\beta}{\sqrt{\kappa_{MC}}}\tilde{x}_{m}}, \\
	\delta \tilde{X}_{L}^{+} & = & \frac{2}{\sqrt{\kappa_{AC}}} \left(\delta \tilde{X}^{+}_{d}+\frac{4Nkg_{a}\alpha_{L}}{\sqrt{\kappa_{AC}}}\tilde{x}_{a}\right), \\
	\delta \tilde{X}_{L}^{-} & = & \frac{2}{\sqrt{\kappa_{AC}}} \left(\delta\tilde{X}_{d}^{-}-\frac{4g_{m}\beta}{\sqrt{\kappa_{MC}}}\tilde{x}_{m}\right). \\
\end{eqnarray*}}
By inspecting these equations we may see that the optomechanical interaction creates phase fluctuations which then perturb the atoms, whilst motion of the atoms modulates the amplitude fluctuations experienced by the micromechanical element.

The optical field at the output may be found by calculating $\delta \baop{a}_{out} = \sqrt{\kappa_{AC}} \delta\baop{a}_{L} - \delta \baop{a}_{in,L}$. Imperfect homodyne detection (efficiency $\eta$) is modelled by applying a standard beamsplitter transformation to this field (Eqn~\eqref{Eqn:OutputField})---which introduces an amount of uncorrelated noise associated with the vacuum fluctuations $\delta \hat{Z}_{v}^{\pm}$---and treating the photodetectors as perfectly efficient. The resulting effective detected field is
\begin{eqnarray*}
\delta \hat{Z}^{+} & = & \sqrt{\eta}\delta \hat{X}_{d}^{+} + \sqrt{1-\eta}\delta\hat{Z}_{v}^{+}, \\
\delta \hat{Z}^{-} & = & \sqrt{\eta}\Br{\delta \hat{X}_{d}^{-}-\frac{4g_{m}\beta}{\sqrt{\kappa_{MC}}}\x_{m}} + \sqrt{1-\eta}\delta \hat{Z}_{v}^{-}.
\end{eqnarray*}

\subsection[\hspace{3cm} Back-action forces]{\textbf{Back-action forces}} \label{Sec:BackAction}
Substituting the above into the equations of motion (Eqn~\eqref{Eqn:MechAC} and Eqn~\eqref{Eqn:MechMT}) gives the time evolution of the mechanical elements, Eqn~\eqref{Eqn:MechACEffNoLoss} and Eqn~\eqref{Eqn:MechMTEffNoLoss}, under the effect of the coupling (Eqn~\eqref{Eqn:CouplingRate}) and the back-action noise
\begin{eqnarray*}
\tilde{F}_{BA,a} & = & 0 ,\\
\tilde{F}_{BA,m} & = & \frac{-4\hbar g_{m}\alpha_{d}}{\kappa_{MC}}\delta\tilde{X}_{d}^{+}. 
\end{eqnarray*}

We briefly discuss the result $\tilde{F}_{BA,a} = 0$. Complete cancellation of the optical back-action onto the atomic motion is an artefact of our model, arising due to neglect of optical loss and near-field atom--atom interactions. In reality, there will be some amount of heating caused by optical loss, near-field interactions between atoms and spontaneous scattering of photons out of the trap beams.

As noted in $\S$~\ref{Sec:EffectiveDynamics}, any optical loss will render the effective oscillator--oscillator interaction asymmetric, and hence non-Hamiltonian. Such losses also introduce vacuum noise on the left-going field which is uncorrelated with that on the right-going, leading to imperfect cancellation of the phase fluctuations appearing in the atomic equation of motion. In the limit that the noise is completely uncorrelated the back-action heating rate is $2\Gamma_{a}c_{a}$.

Even in the absence of uncorrelated noise, diffusion of the atomic centre of mass occurs due to near-field optically-mediated interactions between atoms: for instance, a sideband photon may be emitted by one atom and absorbed by another. These processes do not alter the optical far-field, which our (one-dimensional) model describes, but do lead to back-action heating of the atomic motion. This effect will be small in the far-detuned limit, and scales weakly with the atom number ($\propto N^{1/3}$) \cite{Balykin2000}, and we therefore neglect it.

The heating rate associated with Gordon-Ashkin (GA) diffusion \cite{Gordon1980} is also negligible in the regime discussed. To illustrate this, consider the momentum diffusion coefficient $\mathcal{D}_{p}$ given by \cite{Murr2006}. Since the atoms are trapped near to an antinode of the cavity field there is (to first order) no spatial variation of the electric field amplitude or of the degree of coherence between the atoms' ground and excited states (given by $\expect{\sigma}$, where $\sigma$ is the atomic lowering operator), so the diffusion should be dominated by spontaneous scattering terms. The axial motion of each atom is thus heated at a rate (in the harmonic approximation)
\[
\Gamma_{GA}\bar{n}_{GA} = \frac{\mathcal{D}_{p}}{6m\hbar\omega_{a}} \approx \frac{\omega_{a}\gamma_{e}}{8\modd{\Delta_{t}}},
\]
where the excited state lifetime is $1/2\gamma_{e}$ (on the order of $10$ ns for \isotope{Rb}{87} \cite{Steck2003}), $\Gamma_{GA}$ is the coupling rate to the effective bath and $\bar{n}_{GA}$ is the phonon number characterising this bath. We may easily satisfy $\Gamma_{GA}\bar{n}_{GA} \ll \Gamma_{a}/2$ because we operate in the far-detuned and bad-cavity limits. Furthermore, independent scattering from each atom results in suppression of the centre of mass diffusion coefficient by a factor of $1/\sqrt{N}$. Heating due to Gordon-Ashkin processes may therefore be safely omitted from our model.

Incorporating these imperfections into the model will leave the essential conclusions of this paper unchanged. Qualitatively, the chief difference is that near-ground-cooled `tongue' evident in Fig.~\ref{Fig:Sympathetic} and Fig.~\ref{Fig:CombinedCooling} does not extend to arbitrarily high $c_{a}$ (at low $c_{m}$): at some point the atomic back-action heats the mechanical oscillator out of the ground state. We then expect, as predicted by \cite{Hammerer2010} and \cite{Vogell2013}, that the ground state may be approached by sympathetic cooling if the both the atoms and mechanics operate in the regime where the back-action and thermal (plus zero-point) noises are approximately equal.

\section[\hspace{3cm} Correlators]{Correlators} \label{Sec:Correlators}

If the typical thermal timescale $\hbar / k_{B}T_{B,m}$ is small compared to the mechanical period it is possible to form the correlator \cite{Genes2008a}
\begin{equation*}
\expect{\tilde{F}_{TH}\left(\omega\right)\tilde{F}_{TH}\left(\omega^{\prime}\right)} = 4\pi\hbar\Gamma_{m}\omega_{m}M\left(\bar{n}_{B,m}+\half\right)\delta\left(\omega+\omega^{\prime}\right).
\end{equation*}
$\bar{n}_{B,m}$ is understood to be the mean number of excitations in the oscillator when in thermal equilibrium with its bath.

When forming the product $\expect{\tilde{x}_{m}\fd\tilde{x}_{m}\Br{\omega^{\prime}}}$ the following correlators also arise;
\begin{eqnarray*}
\expect{\tilde{F}_{CB}\left(\omega\right)\tilde{F}_{CB}\left(\omega^{\prime}\right)} & = & 4\pi\hbar\Gamma_{a}Nm\omega_{a}\half\;\delta\Br{\omega+\omega^{\prime}}, \\
\expect{\tilde{F}_{BA,m}\left(\omega\right)\tilde{F}_{BA,m}\left(\omega^{\prime}\right)} & = & 4\pi\hbar\Gamma_{m}M\omega_{m}c_{m}\;\delta\Br{\omega+\omega^{\prime}}, \\
\expect{\tilde{F}_{SN}\left(\omega\right)\tilde{F}_{BA,m}\left(\omega^{\prime}\right)} & = & 4\pi\hbar\Gamma_{m}M\omega_{m}\frac{G}{4}\frac{\omega/\omega_{m}}{1-\I\omega/\Delta\omega_{FB}}\delta\Br{\omega+ \omega^{\prime}}, \\
\expect{\tilde{F}_{SN}\left(\omega\right)\tilde{F}_{SN}\left(\omega^{\prime}\right)} & = &  -4\pi\hbar\Gamma_{m}\omega_{m}M\left(\frac{G^{2}}{4^{2}\eta c_{m}}\right) \times \\
& & \phantom{-}\frac{\omega\omega^{\prime}}{\omega_{m}^{2}}\frac{\delta\left(\omega+\omega^{\prime}\right)}{\left(1-\I\omega/\Delta\omega_{FB}\right)\left(1-\I\omega^{\prime}/\Delta\omega_{FB}\right)}. \\
\end{eqnarray*}
It is important to recall that, for any observables $\hat{A}$ and $\hat{B}$, $\expect{\tilde{A}\fd\tilde{B}\left(\omega^{\prime}\right)} = \expect{\tilde{B}\left(-\omega^{\prime}\right)\tilde{A}\left(-\omega\right)}^{*}$. We note also that the commutation relations ensure that for any vacuum field $\expect{\delta\tilde{Z}^{\pm}\left(\omega\right)\delta\tilde{Z}^{\mp}\left(\omega^{\prime}\right)} = \pm 2\pi \I \delta(\omega + \omega^{\prime})$.

All correlators not obtained from those above vanish.

\section[\hspace{3cm} Useful integrals]{Useful integrals} \label{Sec:Maths}

The following integrals arise in the analytical evaluation of $\expect{\x_{m}^{2}}$ (\cf{} Eqn~\eqref{Eqn:VarianceGeneral});
\begin{eqnarray*}
\frac{\pi}{\Gamma\Omega^{2}} & = & \intinf{\omega} \frac{1}{\left(\omega^{2}-\Omega^{2}\right)^{2}+\Gamma^{2}\omega^{2}}, \\
\frac{\pi}{\Gamma} & = & \intinf{\omega} \frac{\omega^{2}}{\left(\omega^{2}-\Omega^{2}\right)^{2}+\Gamma^{2}\omega^{2}}.
\end{eqnarray*}

\section[\hspace{3cm} Approximations to the mechanical susceptibility]{Approximations to the mechanical susceptibility} \label{Sec:ApproxAnalytical}

As noted in $\S$~\ref{Sec:AtomsAlone}, the primary challenge in analytically determining $\expect{\x_{m}^{2}}$ is finding suitable approximations to
\begin{eqnarray*}
\intinf{\omega} & & \modd{\chi_{m}^{\prime}}^{2},\\
\intinf{\omega} & & \modd{\chi_{m}^{\prime}}^{2}\modd{\chi_{a}}^{2}
\end{eqnarray*}
in the weak and strong coupling regimes.

\begin{figure}[bt]
\centering
\def\svgwidth{1\columnwidth}
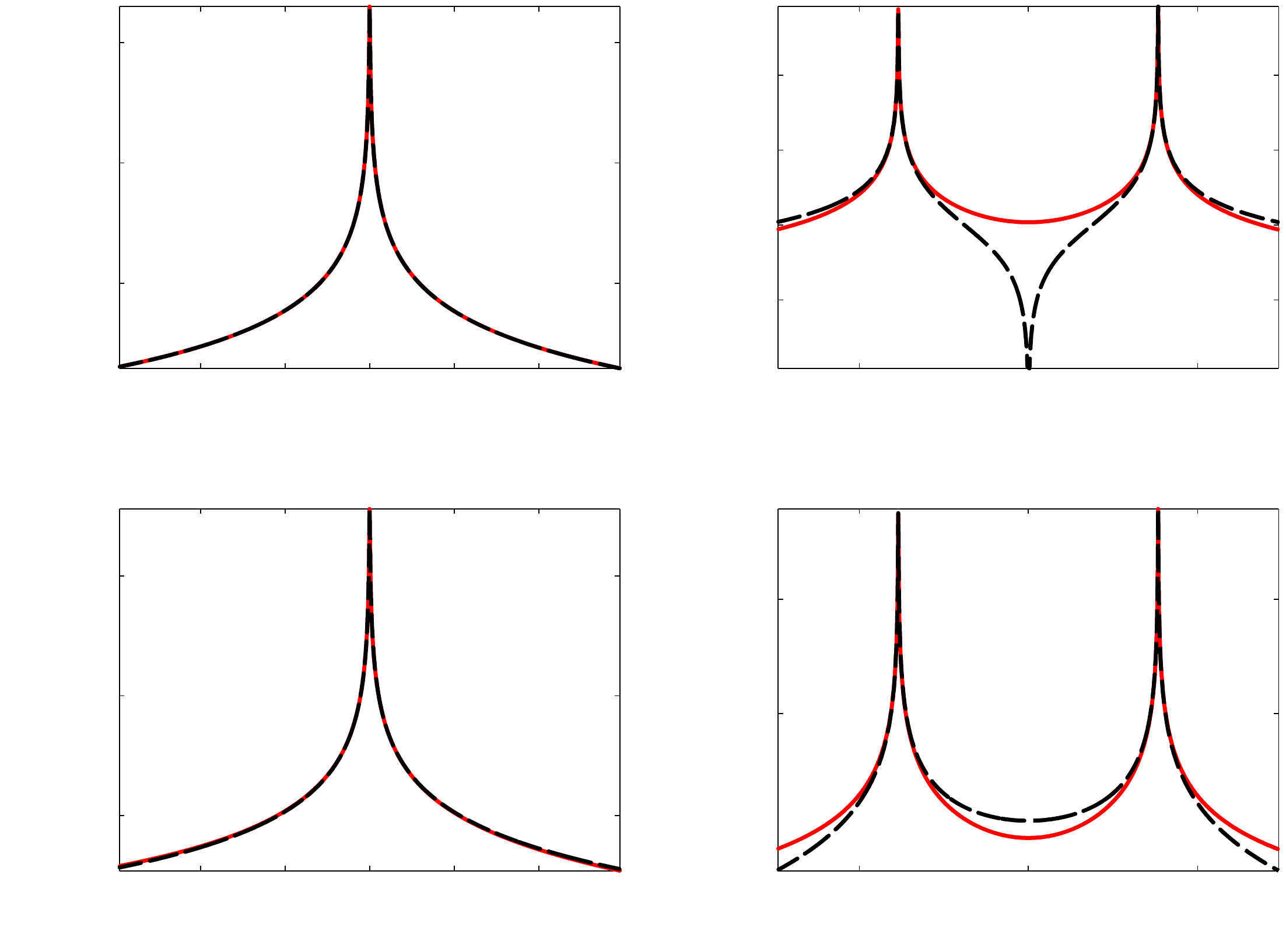
\caption{\label{Fig:SusceptibilityApproximations} Integrands which arise in the calculation of $\expect{\x_{m}^{2}}$ in the strong coupling regime. Exact expressions are given in black (dashed) and approximations in red (solid). The coupling rate satisfies $g/\Omega \sim 10^{-3}$ and the frequency scale is normalised and translated such that $\Omega \rightarrow 0$. A) and B) are calculated with the same parameters as Fig.~\ref{Fig:Sympathetic}~A), whilst C) and D) draw values from Fig.~\ref{Fig:Sympathetic}~C).\newline
A) In the adiabatic regime the mechanical susceptibility is broadened by the interaction with the atom cloud.\newline
B) Excitations arriving at the mechanical oscillator through the atomic motion have an approximately Lorentzian spectrum ($\modd{\chi_{a}}^{2}\modd{\chi_{m}^{\prime}}^{2}$) in the weak coupling limit.\newline
C) Coherent exchange of excitations between the atoms and mechanics results in the formation of hybrid modes. Their interference suppresses the mechanical response at $\Omega$.\newline
D) The noise entering the mechanical mode \via{} coupling to the atoms is sharply peaked near $\Omega$, but suppression of the mechanical susceptibility in this region (due to interference of the normal modes) ensures that $\modd{\chi_{a}}^{2}\modd{\chi_{m}^{\prime}}^{2}$ has only two peaks. Our approximation to this integrand is most accurate in the regions with the largest spectral variance.}
\end{figure}

The derivation of an approximation to the former in the weak coupling limit is given in $\S$~\ref{Sec:AtomsAlone} (plotted in Fig.~\ref{Fig:SusceptibilityApproximations}~A).

In the same limit the latter integral may be evaluated by using the fact that the majority of the spectral variance is concentrated near the resonance frequency $\Omega$. We therefore replace $\modd{\chi_{m}^{\prime}}^{2}\modd{\chi_{a}}^{2}$, a product of two approximately Lorentzian functions, with a single (approximate) Lorentzian, the linewidth of which is chosen to give an accurate fit in the region $\omega \approx \Omega$. We find that
\begin{equation*}
\modd{\chi_{m}^{\prime}}^{2}\modd{\chi_{a}}^{2} \approx \Sq{\frac{1}{NmM\Omega\Gamma_{ser}}}^{2}\frac{1}{\Br{\Omega^{2}-\omega^{2}}^{2}+\omega^{2}\Gamma_{par}^{2}}
\end{equation*}
is a suitable replacement (depicted in Fig.~\ref{Fig:SusceptibilityApproximations}~B). The linewidth of this function is the `parallel sum'
\begin{equation*}
\Gamma_{par} = \Br{\frac{1}{\Gamma_{a}}+\frac{1}{\Gamma_{m}\Br{1+c}}}^{-1}
\end{equation*}
of the atomic and (enhanced) mechanical motion decay rates, rather than the `serial sum' $\Gamma_{ser} = \Sq{\Gamma_{a}+\Gamma_{m}\Br{1+c}}$; this is necessary to obtain the correct behaviour when $\Gamma_{m}\Br{1+c} \sim \Gamma_{a}$. Numerical integration confirms that this replacement faithfully reproduces the true value of $\intinf{\omega} \modd{\chi_{m}^{\prime}}^{2}\modd{\chi_{a}}^{2}$, despite decaying more slowly than the true integrand as $\omega\rightarrow \infty$.

In the regime that $\Omega/2 > g \gg \max\Cu{\Gamma_{m}, \Gamma_{a}}$ the atoms no longer adiabatically follow the motion of the mechanical oscillator; instead, the coherent interaction of the two resonators leads to hybridisation of the mechanical modes. Splitting of the susceptibility $\modd{\chi_{m}^{\prime}}^{2}$ into two peaks, evident in Fig.~\ref{Fig:SusceptibilityApproximations}~C, is a clear signature of hybridisation.

Since the two peaks of the susceptibility account for the majority of the spectral variance we set 
\begin{equation}
\modd{\chi_{m}^{\prime}}^{2} \approx \modd{\chi_{+}}^{2} + \modd{\chi_{-}}^{2}
\label{Eqn:IntegralApproxDoublePeak}
\end{equation}
where the symmetric ($+$) and antisymmetric ($-$) modes have susceptibilities
\begin{equation}
\chi_{\pm}^{-1} = M_{\pm}\Sq{\Omega^{2}\Br{1\pm\frac{-2g}{\Omega}}-\omega^{2}-\I\omega\Gamma_{N}}.
\label{Eqn:PMSusceptibility}
\end{equation}
Note that when $g \ll \Omega/2$ the splitting between the symmetric and antisymmetric peaks is $2g$. The linewidth $\Gamma_{N}$ is the mean of $\Gamma_{m}$ and $\Gamma_{a}$, and the effective masses $M_{\pm} = 2M\Br{1-g^{2}/\Omega^{2}}$ may be obtained by considering the zero-frequency behaviour of $\modd{\chi_{m}^{\prime}}^{2}$. It is usually possible to ignore the strong suppression of the susceptibility at $\omega \approx \Omega$ (\cf{} Fig.~\ref{Fig:SusceptibilityApproximations}~C), due to interference of the normal modes, without significantly altering the value of the integral.

Noise acting on the atomic local mode appears in $S\sb{x_{m}x_{m}}\Sq{\omega}$ as a sharp peak at $\omega \approx \Omega$; fortunately, the aforementioned interference of the normal modes ensures that the product $\modd{\chi_{a}}^{2}\modd{\chi_{m}^{\prime}}^{2}$ remains dominated by the twin-peaked susceptibility, as in Fig.~\ref{Fig:SusceptibilityApproximations}~D. Thus we again note that the majority of the spectral variance lies close to the maxima, motivating the substitution
\begin{equation}
\modd{\chi_{a}}^{2}\modd{\chi_{m}^{\prime}}^{2} \approx \frac{\modd{\chi_{+}}^{2} + \modd{\chi_{-}}^{2}}{\Br{2Nm\Omega g}^{2}}.
\label{Eqn:IntegralApproxProduct}
\end{equation}
The scaling factor $\Br{2Nm\Omega g}^{-2}$ accounts for the different masses of the normal and local modes. This approximate integrand yields quite accurate results, as confirmed by numerical integration.

Combining Eqn~\eqref{Eqn:IntegralApproxDoublePeak} and Eqn~\eqref{Eqn:IntegralApproxProduct}, we find that the variance is given by
\begin{equation*}
\frac{\expect{\x_{m}^{2}}}{x_{zp,m}^{2}} = \half\Sq{\frac{\Gamma_{m}}{\Gamma_{N}}\Br{\bar{n}_{B,m} + c_{m}} + 1}{\Br{1-\frac{g^{2}}{\Omega^{2}}}^{-2}\Br{1-\frac{4g^{2}}{\Omega^{2}}}}^{-1}
\end{equation*}
as previously stated in $\S$~\ref{Sec:AtomsAlone}.

\begin{thebibliography}{10}
\expandafter\ifx\csname url\endcsname\relax
  \def\url#1{{\tt #1}}\fi
\expandafter\ifx\csname urlprefix\endcsname\relax\def\urlprefix{URL }\fi
\providecommand{\eprint}[2][]{\url{#2}}

\bibitem{Treutlein2012}
Treutlein P, Genes C, Hammerer K, Poggio M and Rabl P 2012 (\textit{Preprint} \eprint{arXiv:1210.4151v1})

\bibitem{Chen2013}
Chen Y 2013 {\em Journal of Physics B: Atomic, Molecular and Optical Physics\/}
  {\bf 46} 104001

\bibitem{Giovannetti2004}
Giovannetti V, Lloyd S and Maccone L 2004 {\em Science\/} {\bf 306} 1330--1336

\bibitem{Milburn2011}
Milburn G~J and Woolley M~J 2011 {\em Acta Physica Slovaca\/} {\bf 61} 483 --
  601

\bibitem{Serafini2012}
Serafini A 2012 {\em International Scholarly Research Network Optics\/} {\bf
  2012} 275016

\bibitem{Tsukanov2011}
Tsukanov A~V 2011 {\em Mikroelektronika (Russian Microelectronics)\/} {\bf 40}
  359--369

\bibitem{Muschik2011}
Muschik C~A, Krauter H, Hammerer K and Polzik E~S 2011 {\em Quantum Information
  Processing\/} {\bf 10} 839--863

\bibitem{Stannigel2012}
Stannigel K, Komar P, Habraken S~J~M, Bennett S~D, Lukin M~D, Zoller P and Rabl
  P 2012 {\em Physical Review Letters\/} {\bf 109} 013603

\bibitem{Kleckner2008}
Kleckner D, Pikovski I, Jeffrey E, Ament L, Eliel E, van~den Brink J and
  Bouwmeester D 2008 {\em New Journal of Physics\/} {\bf 10} 095020

\bibitem{Rocheleau2010}
Rocheleau T, Ndukum T, Macklin C, Hertzberg J~B, Clerk A~A and Schwab K~C 2010
  {\em Nature\/} {\bf 463} 72--75

\bibitem{OConnell2010}
O'Connell A~D, Hofheinz M, Ansmann M, Bialczak R~C, Lenander M, Lucero E,
  Neeley M, Sank D, Wang H, Weides M, Wenner J, Martinis J~M and Cleland A~N
  2010 {\em Nature\/} {\bf 464} 697--703

\bibitem{Teufel2011}
Teufel J~D, Donner T, Li D, Harlow J~W, Allman M~S, Cicak K, Sirois A~J,
  Whittaker J~D, Lehnert K~W and Simmonds R~W 2011 {\em Nature\/} {\bf 475}
  359--363

\bibitem{Chan2011}
Chan J, {Mayer Alegre} T~P, Safavi-Naeini A~H, Hill J~T, Krause A, Gr\"{o}blacher
  S, Aspelmeyer M and Painter O 2011 {\em Nature\/} {\bf 478} 89--92

\bibitem{Hammerer2010}
Hammerer K, Stannigel K, Genes C, Zoller P, Treutlein P, Camerer S, Hunger D
  and H\"{a}nsch T~W 2010 {\em Physical Review A\/} {\bf 82} 021803(R)

\bibitem{Camerer2011}
Camerer S, Korppi M, J\"{o}ckel A, Hunger D, H\"{a}nsch T~W and Treutlein P 2011 {\em
  Physical Review Letters\/} {\bf 107} 223001

\bibitem{Vogell2013}
Vogell B, Stannigel K, Zoller P, Hammerer K, Rakher M~T, Korppi M, J\"{o}ckel A
  and Treutlein P 2013 {\em Physical Review A\/} {\bf 87} 023816

\bibitem{Mancini1998}
Mancini S, Vitali D and Tombesi P 1998 {\em Physical Review Letters\/} {\bf 80}
  688--691

\bibitem{Courty2001}
Courty J, Heidmann A and Pinard M 2001 {\em The European Physical Journal D\/}
  {\bf 17} 399--408

\bibitem{Vitali2002}
Vitali D, Mancini S, Ribichini L and Tombesi P 2002 {\em Physical Review A\/}
  {\bf 65} 063803

\bibitem{Hopkins2003}
Hopkins A, Jacobs K, Habib S and Schwab K 2003 {\em Physical Review B\/} {\bf
  68} 235328

\bibitem{Genes2008a}
Genes C, Vitali D, Tombesi P, Gigan S and Aspelmeyer M 2008 {\em Physical
  Review A\/} {\bf 77} 033804

\bibitem{Doherty2012}
Doherty A~C, Szorkovszky A, Harris G~I and Bowen W~P 2012 {\em Philosophical
  Transactions of the Royal Society A: Mathematical, Physical and Engineering
  Sciences\/} {\bf 370} 5338--5353

\bibitem{Jacobs2014}
Jacobs K 2014 (\textit{Preprint}
  \eprint{arXiv:1304.0819v2})

\bibitem{Lloyd2000}
Lloyd S 2000 {\em Physical Review A\/} {\bf 62} 022108

\bibitem{Hamerly2012}
Hamerly R and Mabuchi H 2012 {\em Physical Review Letters\/} {\bf 109} 173602

\bibitem{Hammerer2009EPR}
Hammerer K, Aspelmeyer M, Polzik E~S and Zoller P 2009 {\em Physical Review
  Letters\/} {\bf 102} 020501

\bibitem{Kimble2008}
Kimble H~J 2008 {\em Nature\/} {\bf 453} 1023--1030

\bibitem{Grimm2000}
Grimm R, Weidem\"{u}ller M and Ovchinnikov Y 2000 {\em Advances in Atomic,
  Molecular and Optical Physics\/} {\bf 95} 95--170

\bibitem{Dalibard1985}
Dalibard J and Cohen-Tannoudji C 1985 {\em Journal of the Optical Society of
  America B\/} {\bf 2} 1707--1720

\bibitem{WallsMilburn2008}
Milburn G~J 2008 {\em {Quantum Optics}\/} 2nd ed vol~- (Berlin: Springer) ISBN
  978-3-540-28573-1

\bibitem{Kippenberg2007}
Kippenberg T~J and Vahala K~J 2007 {\em Optics Express\/} {\bf 15} 74--77

\bibitem{KippenbergVahala2008}
Kippenberg T~J and Vahala K~J 2008 {\em Science\/} {\bf 321} 1172--1176

\bibitem{Thourhout2010}
Thourhout D~V and Roels J 2010 {\em Nature Photonics\/} {\bf 4} 211--217

\bibitem{Law1995}
Law C~K 1995 {\em Physical Review A\/} {\bf 51} 2537--2541

\bibitem{Gordon1980}
Gordon J~P and Ashkin A 1980 {\em Physical Review A\/} {\bf 21} 1606--1617

\bibitem{GardinerZoller}
Gardner C~J and Zoller P 2004 {\em {Quantum Noise: A Handbook of Markovian and
  Non-Markovian Quantum Stochastic Methods with Applications to Quantum
  Optics}\/} 3rd ed (Berlin: Springer) ISBN 3-540-22301-0

\bibitem{Phillips1998}
Phillips W~D 1998 {\em Reviews of Modern Physics\/} {\bf 70} 721--741

\bibitem{Pinard2000}
Pinard M, Cohadon P~F, Briant T and Heidmann A 2000 {\em Physical Review A\/}
  {\bf 63} 013808

\bibitem{Poggio2007}
Poggio M, Degen C~L, Mamin H~J and Rugar D 2007 {\em Physical Review Letters\/}
  {\bf 99} 017201

\bibitem{Lee2010}
Lee K~H, McRae T~G, Harris G~I, Knittel J and Bowen W~P 2010 {\em Physical
  Review Letters\/} {\bf 104} 123604

\bibitem{Sridaran2011}
Sridaran S and Bhave S~A 2011 {\em Optics Express\/} {\bf 19} 9020--9026

\bibitem{Kleckner2006}
Kleckner D and Bouwmeester D 2006 {\em Nature\/} {\bf 444} 75--78

\bibitem{LIGO2009}
LIGO collaboration 2009 {\em New Journal of Physics\/} {\bf 11} 073032

\bibitem{Courty2000}
Courty J~M, Grassia F and Reynaud S 2000 {Thermal and Quantum Noise in Active
  Systems} {\em Noise, Oscillators and Algebraic Randomness - From Noise in
  Communicaton Systems to Number Theory\/} ed Planat M (Chapelle des Bois:
  Springer) chap~3, pp 71--83 ISBN 3540675728

\bibitem{Aspelmeyer2013review}
Aspelmeyer M, Kippenberg T~J and Marquardt F 2013 (\textit{Preprint} \eprint{arXiv:1303.0733v1})

\bibitem{Schmid2011}
Schmid S, Jensen K~D, Nielsen K~H and Boisen A 2011 {\em Physical Review B\/}
  {\bf 84} 165307

\bibitem{Brawley2012}
Brawley G, Bennett J, Cole R, Bowen W~P, Schmid S and Boisen A 2012 {\em
  Proceedings of OSA QO VI\/}  6--7

\bibitem{Thompson2008}
Thompson J~D, Zwickl B~M, Jayich A~M, Marquardt F, Girvin S~M and Harris J~G~E
  2008 {\em Nature\/} {\bf 452} 06715

\bibitem{Vetsch2012}
Vetsch E, Dawkins S~T, Mitsch R, Reitz D, Schneewei\ss{} P and Rauschenbeutel A
  2012 {\em IEEE Journal of Quantum Electronics\/} {\bf 18} 1763

\bibitem{Blencowe2004}
Blencowe M 2004 {\em Physics Reports\/} {\bf 395} 159 -- 222

\bibitem{Murr2006}
Murr K, Maunz P, Pinkse P~W~H, Puppe T, Schuster I, Vitali D and Rempe G 2006
  {\em Physical Review A\/} {\bf 74} 043412

\bibitem{Kerman2000}
Kerman A J, Vuleti\'{c} V, Chin C, and Chu, S 2000
{\em Physical Review Letters\/} {\bf 84} 439--442

\bibitem{Balykin2000}
Balykin V I, Minogin V G, and Letokhov V S 2000
{\em Reports on Progress in Physics\/} {\bf 63} 1429--1510

\bibitem{Steck2003}
Steck D~A 2010 {Rubidium 87 D Line Data} \url{http://steck.us/alkalidata/}

\end{thebibliography}
\end{document}